\documentclass[twocolumn]{aastex631} 
\usepackage{CLASSY_Preamble}

\usepackage{amssymb}	
\usepackage{longtable}
\usepackage{scalerel}

\mathchardef\mhyphen="2D

\newcommand{\angstrom}{\text{ \normalfont\AA}}

\mathchardef\mhyphen="2D

\def\ly{$\lambda$}
\def\ha{H$\alpha$}

\def\hi{H\,{\sc i}}

\def\nii{N\,{\sc ii}}

\def\sii{S\,{\sc ii}}

\def\Q0059{Q0059--2735}

\def\S2S3{S2S3}

\definecolor{blk}{rgb}{0.0,0.0,0.0}
\definecolor{red}{rgb}{0.75,0.0,0.0}
\definecolor{yel}{rgb}{0.65,0.65,0.0}
\definecolor{grn}{rgb}{0.0,0.75,0.0}
\definecolor{blu}{rgb}{0.0,0.0,0.75}
\definecolor{gry}{rgb}{0.75,0.75,0.75}

\def\nh{\ifmmode n_\mathrm{\scriptscriptstyle H} \else $n_\mathrm{\scriptscriptstyle H}$\fi}
\def\ne{\ifmmode n_\mathrm{\scriptstyle e} \else $n_\mathrm{\scriptstyle e}$\fi}
\def\np{\ifmmode n_\mathrm{\scriptstyle p} \else $n_\mathrm{\scriptstyle p}$\fi}
\def\Te{\ifmmode T_\mathrm{\scriptstyle e} \else $T_\mathrm{\scriptstyle e}$\fi}
\def\Qh{\ifmmode Q_\mathrm{\scriptstyle H} \else $Q_\mathrm{\scriptstyle H}$\fi}
\def\Qhesc{\ifmmode Q_\mathrm{\scriptstyle H, esc} \else $Q_\mathrm{\scriptstyle H, esc}$\fi}
\def\Uh{\ifmmode U_\mathrm{\scriptstyle H} \else $U_\mathrm{\scriptstyle H}$\fi}
\def\Nh{\ifmmode N_\mathrm{\scriptstyle H} \else $N_\mathrm{\scriptstyle H}$\fi}
\def\NSi{\ifmmode N_\mathrm{\scriptstyle si} \else $N_\mathrm{\scriptstyle Si}$\fi}
\def\Uhhp{\ifmmode U_\mathrm{\scriptstyle H,HP} \else $U_\mathrm{\scriptstyle H,HP}$\fi}
\def\Nhhp{\ifmmode N_\mathrm{\scriptstyle H,HP} \else $N_\mathrm{\scriptstyle H,HP}$\fi}
\def\Uhvhp{\ifmmode U_\mathrm{\scriptstyle H,VHP} \else $U_\mathrm{\scriptstyle H,VHP}$\fi}
\def\Nhvhp{\ifmmode N_\mathrm{\scriptstyle H,VHP} \else $N_\mathrm{\scriptstyle H,VHP}$\fi}
\def\Nion{\ifmmode N_\mathrm{\scriptstyle ion} \else $N_\mathrm{\scriptstyle ion}$\fi}
\def\nion{\ifmmode n_\mathrm{\scriptstyle ion} \else $n_\mathrm{\scriptstyle ion}$\fi}

\def\Zsun{\ifmmode {\rm Z}_{\odot} \else Z$_{\odot}$\fi}
\def\Msun{\ifmmode {\rm M}_{\odot} \else M$_{\odot}$\fi}
\def\kms{\ifmmode {\rm km~s}^{-1} \else km~s$^{-1}$\fi}
\def\Lya{\ifmmode {\rm Ly}\alpha \else Ly$\alpha$\fi}
\def\Lyb{\ifmmode {\rm Ly}\beta \else Ly$\beta$\fi}
\def\Lyg{\ifmmode {\rm Ly}\gamma \else Ly$\gamma$\fi}
\def\Lyd{\ifmmode {\rm Ly}\delta \else Ly$\delta$\fi}
\def\neaod{\ifmmode n_\mathrm{\scriptscriptstyle AOD} \else $n_\mathrm{\scriptscriptstyle AOD}$\fi}
\def\necrit{\ifmmode n_\mathrm{\scriptstyle cr} \else $n_\mathrm{\scriptstyle cr}$\fi}
\def\ncr{\ifmmode n_\mathrm{\scriptstyle cr} \else $n_\mathrm{\scriptstyle cr}$\fi}
\def\nepi{\ifmmode n_\mathrm{\scriptscriptstyle PI} \else $n_\mathrm{\scriptscriptstyle PI}$\fi}
\def\gtorder{\mathrel{\raise.3ex\hbox{$>$}\mkern-14mu\lower0.6ex\hbox{$\sim$}}}
\def\ltorder{\mathrel{\raise.3ex\hbox{$<$}\mkern-14mu\lower0.6ex\hbox{$\sim$}}}

\def\vro{\ifmmode v_\mathrm{\scriptscriptstyle 1, \scriptstyle r} \else $v_\mathrm{\scriptscriptstyle 1, \scriptstyle r}$\fi}
\def\vrc{\ifmmode v_\mathrm{\scriptscriptstyle 2, \scriptstyle r} \else $v_\mathrm{\scriptscriptstyle 2, \scriptstyle r}$\fi}
\def\vzo{\ifmmode v_\mathrm{\scriptscriptstyle 1, \scriptstyle z} \else $v_\mathrm{\scriptscriptstyle 1, \scriptstyle z}$\fi}
\def\vzc{\ifmmode v_\mathrm{\scriptscriptstyle 2, \scriptstyle z} \else $v_\mathrm{\scriptscriptstyle 2, \scriptstyle z}$\fi}

\newcommand{\Vout}{$V_\text{out}$}

\newcommand{\Vcir}{$V_\text{cir}$}

\newcommand{\Mdot}{$\dot{M}_\text{out}$}
\newcommand{\Pdot}{$\dot{P}_\text{out}$}
\newcommand{\PdotStar}{$\dot{P}_{\star}$}

\newcommand{\Edot}{$\dot{E}_\text{out}$}
\newcommand{\EdotStar}{$\dot{E}_{\star}$}

\newcommand{\Wout}{$W_\text{out}$}

\newcommand{\NHcloud}{$N_\text{H,cl}$}
\newcommand{\Rcloud}{$R_\text{cl}$}
\newcommand{\Mcloud}{$M_\text{cl}$}

\newcommand{\PI}{Paper~I}
\newcommand{\PII}{Paper~II}
\newcommand{\PIII}{Paper~III}

\defcitealias{Berg22}{\PI}
\defcitealias{James22}{\PII}
\defcitealias{Xu2022}{\PIII}

\begin{document}

\submitjournal{AASJournal ApJ}
\shortauthors{Xu et al.}
\shorttitle{}

\title{What are the Radial Distributions of Density, Outflow Rates, and Cloud Structures in the M 82 Wind?}

\author[0000-0002-9217-7051]{Xinfeng Xu}
\affiliation{Center for Astrophysical Sciences, Department of Physics \& Astronomy, Johns Hopkins University, Baltimore, MD 21218, USA}

\author[0000-0001-6670-6370]{Timothy Heckman}
\affiliation{Center for Astrophysical Sciences, Department of Physics \& Astronomy, Johns Hopkins University, Baltimore, MD 21218, USA}

\author[0000-0002-9948-1646]{Michitoshi Yoshida}
\affiliation{National Astronomical Observatory of Japan, Osawa, Mitaka, Tokyo 181-8588, Japan}

\author[0000-0002-6586-4446]{Alaina Henry}
\affiliation{Center for Astrophysical Sciences, Department of Physics \& Astronomy, Johns Hopkins University, Baltimore, MD 21218, USA}
\affiliation{Space Telescope Science Institute, 3700 San Martin Drive, Baltimore, MD 21218, USA}

\author[0000-0001-9490-3582]{Youichi Ohyama}
\affiliation{Institute of Astronomy and Astrophysics, Academia Sinica, 11F of Astronomy-Mathematics Building, No.1, Sec. 4, Roosevelt Rd, Taipei 10617, Taiwan, R.O.C.}

\correspondingauthor{Xinfeng Xu} 
\email{xinfeng@jhu.edu}


\begin{abstract}

Galactic winds play essential roles in the evolution of galaxies through the feedback they provide. Despite intensive studies of winds, the radial distributions of their properties and feedback are rarely observable. Here we present such measurements for the prototypical starburst galaxy, M 82, based on observations by Subaru telescope. We determine the radial distribution of outflow densities (\ne) from the spatially-resolved [\sii] \ly\ly 6717, 6731 emission-lines. We find \ne\ drops from 200 to 40 cm$^{-3}$ with radius ($r$) between 0.5 and 2.2 kpc with a best-fit power-law index of $r^{-1.2}$. Combined with resolved \ha\ lines, we derive mass, momentum, and energy outflow rates, which drop quite slowly (almost unchanged within error bars) over this range of $r$. This suggests that the galactic wind in M 82 can carry mass, momentum, and energy from the central regions to a few kpc with minimal losses. We further derive outflow cloud properties, including size and column densities. The clouds we measure have pressures and densities that are too high to match those from recent theoretical models and numerical simulations of winds. By comparing with a sample of outflows in local star-forming galaxies studied with UV absorption-lines, the above-derived properties for M 82 outflows match well with the published scaling relationships. These matches suggest that the ionized gas clouds traced in emission and absorption are strongly related. Our measurements motivate future spatially resolved studies of galactic winds, which is the only way to map the structure of their feedback effects.


\end{abstract} 



\section{Introduction} 
\label{sec:intro}

Galactic winds, which are driven by energy and momentum supplied by star-formation (SF) or active galactic nuclei (AGNs), play an essential role in the evolution of galaxies \citep[e.g.,][]{Chevalier85, Silk98}. They are responsible for various feedback effects, including regulating SF in galaxies, enriching the intergalactic and circumgalactic medium (IGM and CGM) with metals, and reducing the baryons in galactic discs to solve the ``overcooling problem'' \citep[see reviews in][and references therein]{Naab17, Donahue22, Heckman23}.  

The current understanding of starburst-driven galactic winds is that the energy source is provided by stellar winds and core-collapse supernovae from the population of massive stars \citep{Chevalier85}. These ejecta are thermalized in shocks to produce a very hot (up to $\sim 10^8$ K) fluid which expands outward to form a very fast (up to 3000 km/s) and tenuous wind. This wind fluid interacts with ambient gas clouds, accelerating them outward at velocities of $10^2$ to $10^3$ km/s. These outflowing clouds span a wide range of phases, including hot (few million K), warm ionized ($10^4$ K), neutral atomic, and molecular gas \citep[e.g.,][]{Leroy15}. These outflowing clouds can be readily observed in both emission and absorption \citep[see reviews in][]{Heckman17a, Veilleux20}.

Constraining the impact of galactic winds requires estimating the total mass/energy/momentum that outflows carry. In principle, these can be measured for the different gas phases listed above. These quantities are commonly characterized as outflow rates, i.e., mass/energy/momentum carried by the outflowing material per unit time. In SF galaxies, which are the focus of this paper, the most extensive such studies refer to the warm ionized gas, from UV and optical absorption lines \citep[e.g.,][]{Heckman00,Pettini00,Rupke02, Martin05,Martin06,Grimes09,Weiner09,Rubin10,Steidel10,Martin12,Bordoloi14,Rubin14,Heckman15, Heckman16,Chisholm16a,Chisholm16b,Chisholm17,Sugahara17,Chisholm18, WangWC22, Xu22a}, and from optical emission lines \citep[e.g.,][]{Heckman90, Lehnert96, Newman12a, Newman12b, Rupke13, Wood15, Davies19, Freeman19, Perna19, Rupke19, Swinbank19, Burchett21, Zabl21, Avery22, Marasco22}. 




Nearly all of these previous studies only derived the global outflow rates integrated over the observing aperture, except in a few cases \citep[e.g.,][]{Leroy15, Burchett21}. This is mainly due to the lack of spatially-resolved observations of well-extended outflows in nearby galaxies. Nonetheless, it has been long recognized that outflows in SF galaxies are the collections of gas with a large range of radius, velocity, and gas phase \citep[e.g.,][]{Heckman17a,Veilleux20}. Thus, it is critical to map out the wind properties and outflow rates.

Furthermore, the radial distribution of the structures of the outflowing clouds, including their volume filling factors (FF), densities, column densities (\NHcloud), and sizes (\Rcloud) are largely unknown \citep[but see][]{Xu23}. However, these parameters are not only essential in quantifying feedback effects, but also vital to constraining sub-grid physics in theoretical models and numerical simulations of winds. For the latter, recent studies show that the \Rcloud\ and \Mcloud\ are the key parameters to determine if outflowing clouds can survive long enough to be accelerated by the hot wind fluid \citep[e.g.,][]{Gronke20, Li20, Sparre20, Fielding22}. However, there currently exist no constraints on the radial distribution of these parameters from observations. 




In this paper, we seek to shed light on the radial distribution of outflow rates and cloud properties based on the wealth data on M 82, which is the most intensively studied nearby starburst galaxy. M 82 hosts the best-observed wind in any galaxy, exhibiting clear biconical outflowing multi-phase gas out to distances of at least a few kpc. Detailed studies of the starburst activity and the multi-phase wind are made possible by its proximity \citep[distance $\sim$ 3.89 Mpc,][]{Sakai99}, and are summarized in \cite{Heckman17a}. Combining archival results with new Subaru imaging and resolved spectroscopic data, we aim to tackle various key problems that have not been well-studied previously:
\begin{enumerate}
    \item What are the radial distributions of the mass, momentum, and energy outflow rates? Do the distributions imply that winds can supply sufficient mass, momentum, and energy onto scales large enough to impact the CGM?
    
    \item What are the properties of outflowing clouds at different radii? Are they consistent with the cloud-survival criteria in current theoretical models and numerical simulations?

    \item How are outflow rates connected between different phases? What are the overall combined feedback effects of these phases?

\end{enumerate}

The structure of this paper is as follows. In Section \ref{sec:obs}, we introduce the observations and data. Then we describe how to calculate outflow density, rates, and cloud properties in Section \ref{sec:analysis}, where we also present the radial profiles of these derived parameters. In Section \ref{sec:discuss}, we discuss and compare our results with measurements of outflow rates estimated for other gas phases. We also place the M 82 outflow in the context of a large sample of local SF galaxies and compare our results to theoretical models and numerical simulations. We conclude the paper in Section \ref{sec:conclusion}. Throughout the paper, we adopt a distance to M 82 of 3.89 Mpc \citep{Sakai99}, which leads to 18.9 pc/\arcsec.

\section{Observations and Data}
\label{sec:obs}

\subsection{Imaging Data}
\label{sec:imagedata}
\begin{figure}
\center
	\includegraphics[page = 1, angle=0,trim={0.1cm 0.1cm 0.1cm 0.1cm},clip=true,width=1.0\linewidth,keepaspectratio]{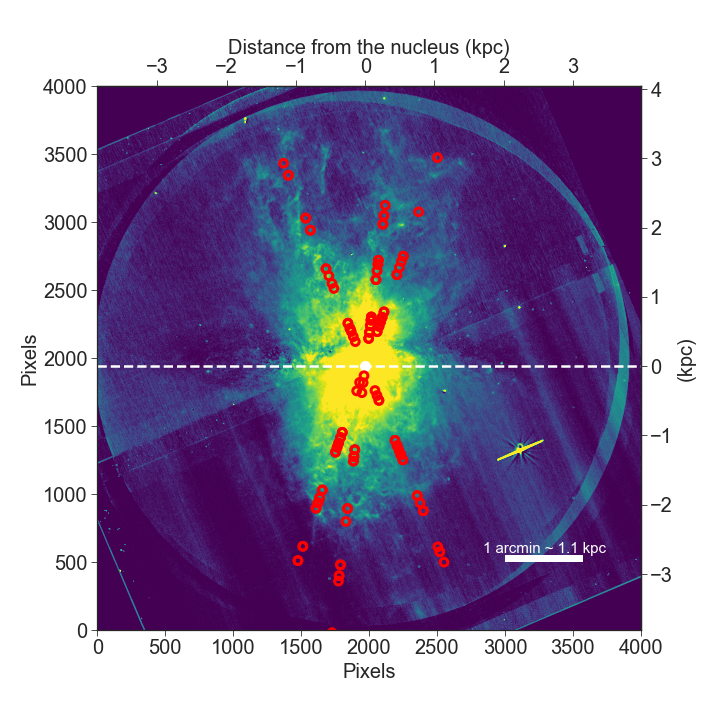}

\caption{\normalfont{Subaru/FOCAS image of \ha\ +[\nii] emission lines (with continuum already subtracted, see Section \ref{sec:imagedata}). The galaxy disk (white dashed line) has been rotated to be parallel with the x-axis, so the bipolar galactic outflow is perpendicular to the galactic disk (north-west side is up). We overlay the position of FOCAS spectroscopic observations described in \cite{Yoshida19} as red circles (Section \ref{sec:specdata}).} }
\label{fig:image}
\end{figure}

Optical imaging observations of M 82 were conducted with Faint Object Camera And Spectrograph \citep[FOCAS,][]{Kashikawa02} on the Subaru Telescope \citep{Kaifu00} on February 2000. These images are published in \cite{Ohyama02}. Two narrow-band filters are used, including the N658 filter that has covered the emission lines of \ha\ \ly 6563 and [\nii] \ly\ly 6548, 6583; and the N642 filter to detect the adjacent continuum level. Their total exposure times are 600s and 360s, respectively, split into 120s sub-exposures to avoid saturation. The seeing was 0.7\arcsec\ to 0.8\arcsec\ during the observations. The data were reduced by pipelines in IDL and IRAF \citep{Yoshida00}. In this paper, we are interested in the \ha\ emission lines. Thus, we subtract the scaled N642 image from the N658 image to get a pure \ha\ +[\nii] emission-line map following the same methodology in \cite{Ohyama02}. The result is shown in Figure \ref{fig:image}.

\subsection {Spectroscopic Data}
\label{sec:specdata}
M 82 was then observed by FOCAS with the spectropolarimetric mode on January 2013. The detailed observations and data reductions are presented in \cite{Yoshida19}. For a summary, the observations adopt a slit mask of eight 0.8\arcsec\ $\times$ 20.6\arcsec\ slitlets at 23.7\arcsec\ intervals and a VPH grism with 665 grooves mm$^{-1}$. These result in a spectral resolution of $R \sim$ 1700 and a central wavelength of 6500\angstrom, which covered the important emission lines for our analyses, including \ha\ and [\sii] \ly\ly 6717, 6731 doublet. The slits are spatially distributed in three position angles (PAs) and at $\sim$ 0 -- 3 kpc from the nucleus. We draw the slit locations as the red circles in Figure \ref{fig:image}. The total science exposure times are 24,000s. The data were reduced using standard CCD reduction pipelines in IRAF in \cite{Yoshida19}. 


\section{Analysis and Results}
\label{sec:analysis}

We start with measuring the radial distributions of the outflow's electron density and width in Sections \ref{sec:density} and \ref{sec:width}, respectively. These values are then used to calculate the radial distribution of outflow rates in Section \ref{sec:rate}. In \citet{Yoshida19} the emission-lines are separated into two components given their spectropolarimetric observations: a polarized component due to scattering of the emission from the central starburst by dust in the outflow, and the total light, which is dominated by the intrinsic emission from the outflowing ionized gas. In this paper, we only use the latter. 

Since our main tracers are \ha\ and [\sii], our calculations in this paper only represent the properties of the warm-ionized phase of outflows (T $\sim$ 10$^{4}$ K). As described in previous publications for M 82 \citep[e.g.,][]{Strickland07, Strickland09}, this warm phase is presumably immersed in and interacting with a dilute volume-filling hot wind fluid (T $\sim$ few $\times$ 10$^{7}$ -- 10$^{8}$ K). Hereafter, we distinguish these two components by referring to the former as (warm) outflows, while the latter as (hot) winds.




\subsection{Radial Distribution of Outflow Densities}
\label{sec:density}
\begin{figure}
\center
	\includegraphics[page = 1, angle=0,trim={0.1cm 0.5cm 0.3cm 0.5cm},clip=true,width=1.0\linewidth,keepaspectratio]{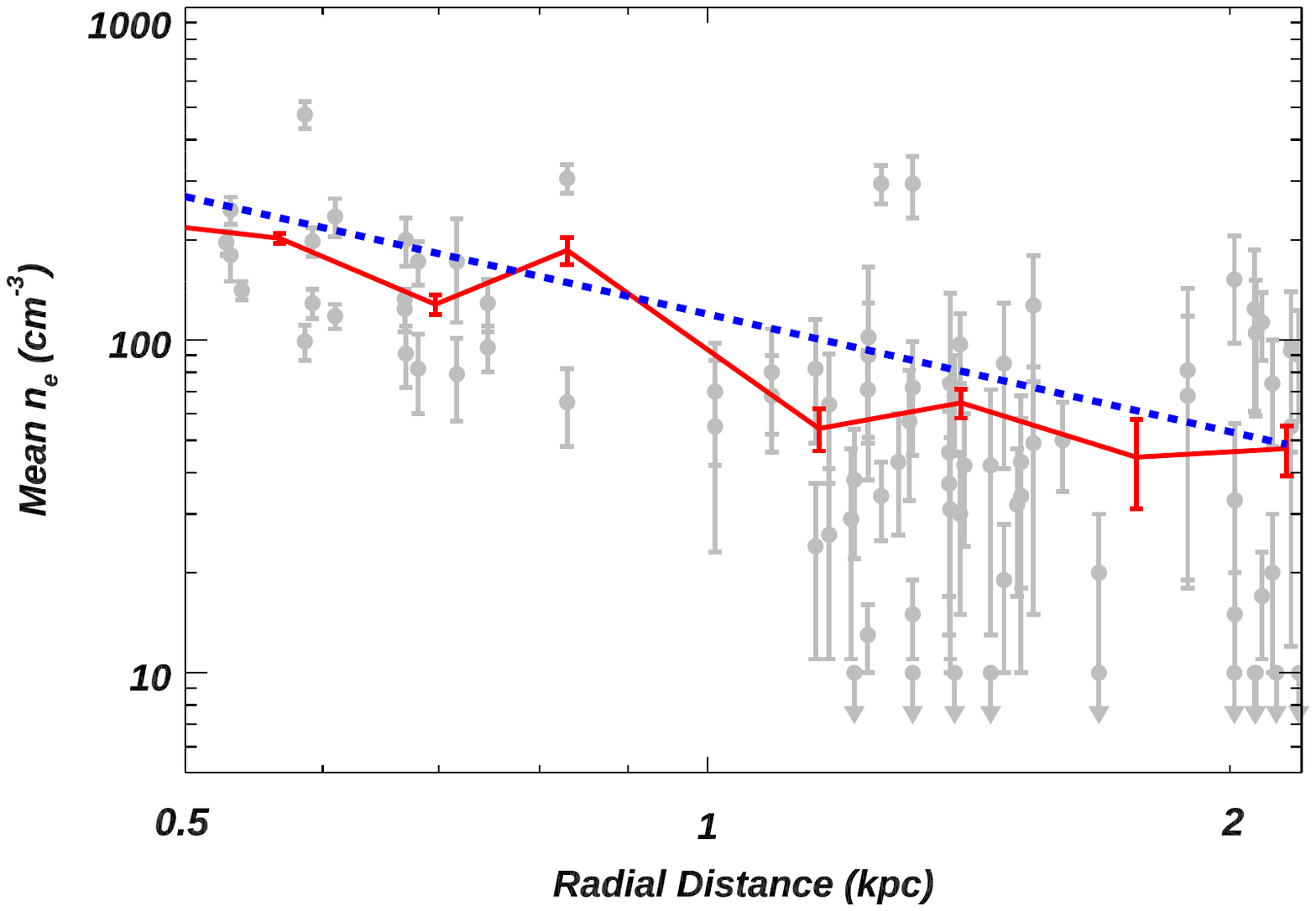}

\caption{\normalfont{Radial Distribution of electron number density (\ne) for M 82's galactic outflows. The gray symbols are the measurements from [\sii] doublet reported in \cite{Yoshida19} at different slit positions (Figure \ref{fig:image}). Their upper limits are shown as arrows. The red curve is the binned \ne\ at different radii considering the gray data points. We have adopted survival analyses to include the upper limits. We also show the best-fit power-law as the blue dashed line.} See section \ref{sec:density}).} 
\label{fig:OutflowDensity}
\end{figure}

\cite{Yoshida19} have already derived the electron density (\ne) from the intensity ratio of [\sii] \ly6731/\ly6717 for each slit location, assuming a typical ionized gas temperature of 10$^{4}$ K \citep{Osterbrock06}. These measurements represent a luminosity-weighted mean density at certain radius. However, these values show moderate scatter (due to intrinsic variations or low S/N) and also contain upper limits (see the gray symbols in Figure \ref{fig:OutflowDensity}). Thus, to account for the scatters, we bin their values to get robust estimates of \ne\ at different radial distances to the galactic center (i.e., $r$) as follows. We first split the measurements into radial bins given $r_{min}$ = 0.5 kpc, $r_{max}$ = 2.5 kpc, and log($\Delta r$) = 0.1 dex. The cut of $r_{min}$ is because the central region of M 82 is dominated by starburst and does not show clear features of outflows \citep[][]{Shopbell98,Westmoquette13}. The cut of $r_{max}$ is due to fewer reliable measurements at larger $r$ reported in \cite{Yoshida19}. This is because the [\sii]-based density measurement reaches its low-density limit at $n_e \sim 10$ cm$^{-3}$ \citep{Osterbrock06}, so direct measurements of \ne\ at larger radii (with lower densities) are not possible. Thus, we adopt survival analysis (which incorporates these upper limits\footnote{We use the \textit{survfit} packge in R-language \citep{R2021}}) to calculate the average \ne\ in each radial bin. The results are shown as the red curve in Figure \ref{fig:OutflowDensity}. We find that \ne\ declines from $\sim$ 200 cm$^{-3}$ at $r$ = 0.5 kpc to $\sim$ 40 cm$^{-3}$ at $r$ = 2.2 kpc. We have fit the data to a power-law and find that $\ne(r) = 100 \times (\frac{r}{1165 pc})^{-1.17}$ cm$^{-3}$ (see the blue dashed line in Figure \ref{fig:OutflowDensity}).


\begin{figure*}
\center
	\includegraphics[angle=0,trim={0.0cm 0.0cm 0.0cm 0.5cm},clip=true,width=0.45\linewidth,keepaspectratio,page=2]{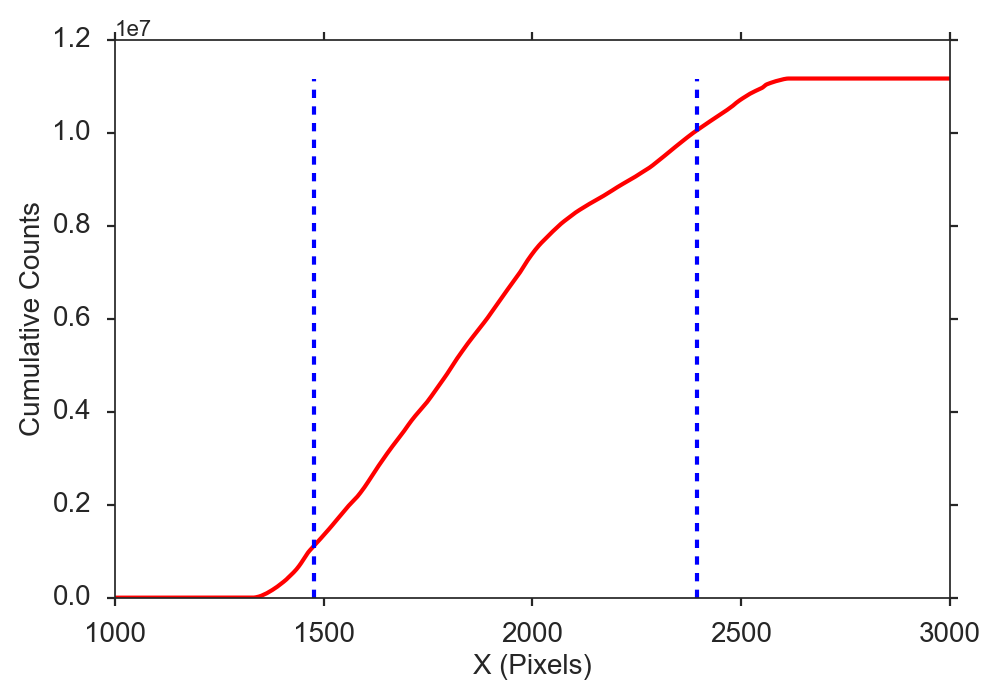}
	\includegraphics[angle=0,trim={0.0cm 0.0cm 0.0cm 0.5cm},clip=true,width=0.45\linewidth,keepaspectratio,page=3]{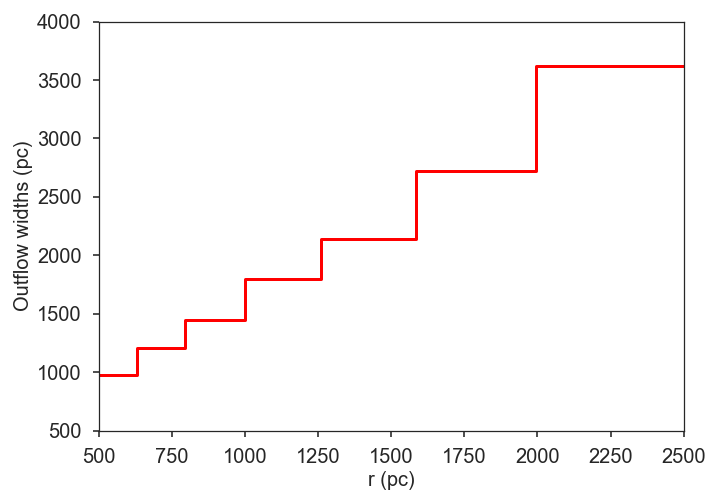}
	
\caption{\normalfont{\textbf{Left:} The cumulative counts of \ha\ from Subaru FOCAS image (see Figure \ref{fig:image} and Section \ref{sec:width}). The two blue dashed lines represent the locations where the cumulative counts equal 10\% and 90\% of the total counts, separately. We define the outflow width as the region between the blue dashed lines, which is 1.8 kpc for this radial bin. \textbf{Right:} Outflow width distribution among different radial distances to the galactic center of M 82 ($r$).} }
\label{fig:OutflowWidth}
\end{figure*}

\subsection{Radial Distribution of the Lateral Outflow Widths}
\label{sec:width}
As shown in Figure \ref{fig:image}, the galactic wind exhibits a bi-conical structure in \ha, which is roughly perpendicular to the galaxy disk (rotated to be the x-axis, white dashed line). To better quantify the regions occupied by the \ha\ outflows, we estimate the lateral width of the outflow (\Wout) at different radial scales as follows.

To remove the background, we first calculate the average counts in blank regions of the image and subtract them from the entire image. Then, for each of the radial bins adopted in \ref{sec:density}, we draw the cumulative \ha\ counts (F$_\text{cum}$) as a function of increasing x. An example is shown in the left panel of Figure \ref{fig:OutflowWidth}. We then define the lateral width of the \ha\ outflow (i.e., \Wout (\ha)) as the region between F$_\text{cum}$ = 10\% and F$_\text{cum}$ = 90\%. The resulting radial distribution of \Wout (\ha) is shown in the right panel. We find that \Wout (\ha) $\sim$ 1.5 $\times$ $r$. If we assume the \ha\ outflow structure is cone-like, this corresponds to an opening angle of 73.7$^{\circ}$ or a solid angle of 0.4 $\pi$ ster (per cone).

\begin{figure*}
\center
	\includegraphics[angle=0,trim={0.0cm 3.5cm 2.0cm 2.1cm},clip=true,width=0.45\linewidth,keepaspectratio]{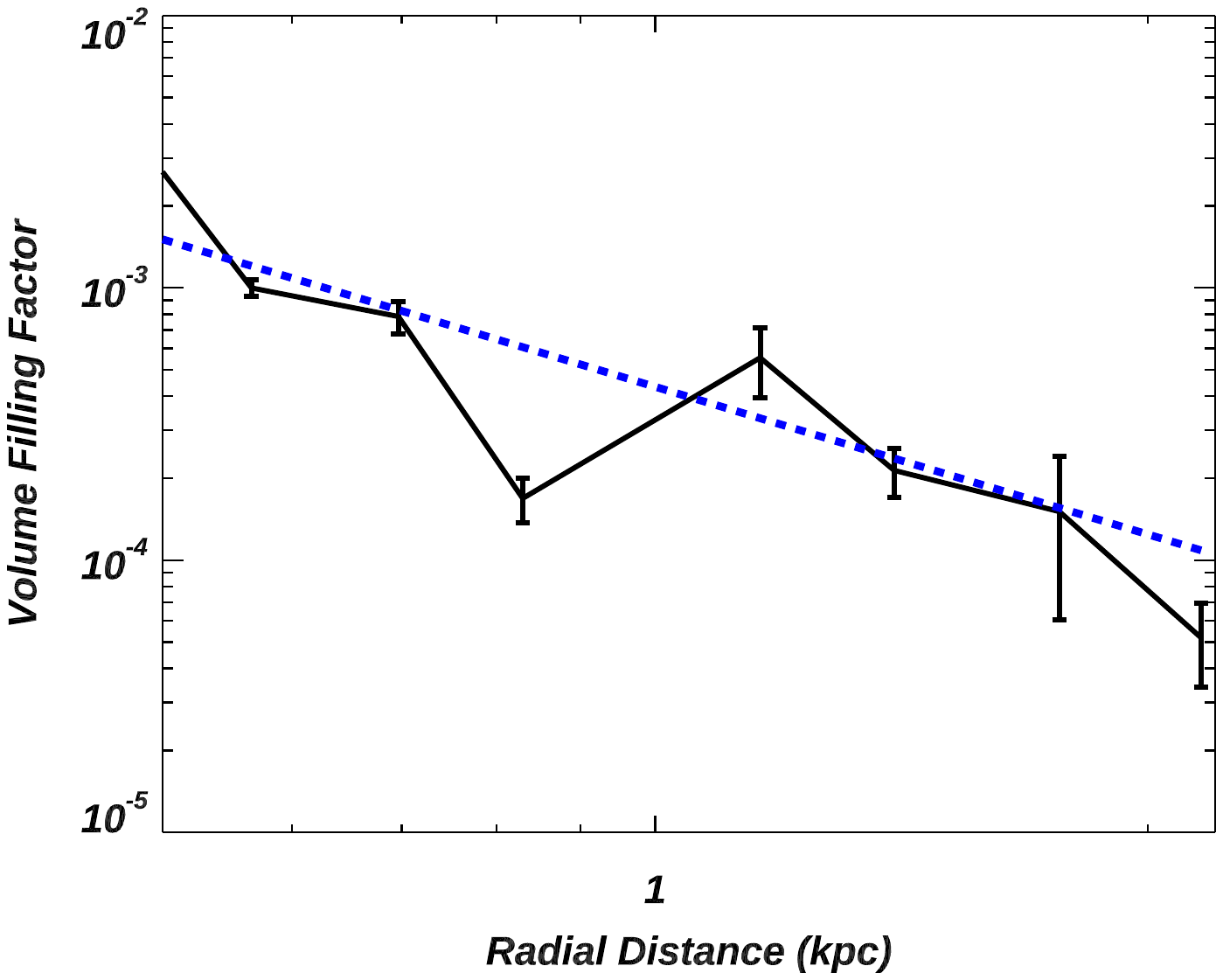}
	\includegraphics[angle=0,trim={0.0cm 3.5cm 0.0cm 2.1cm},clip=true,width=0.48\linewidth,keepaspectratio, page=1]{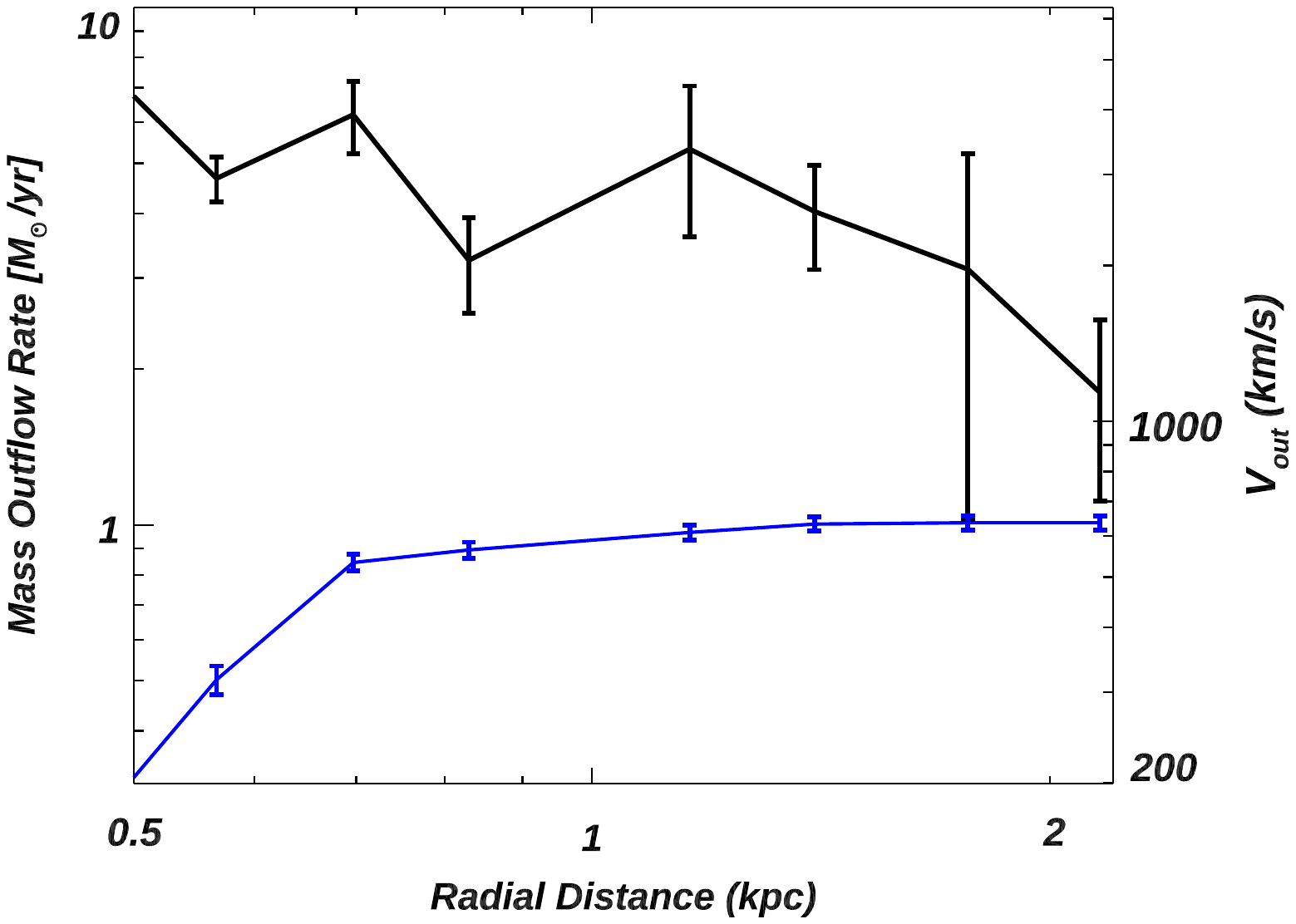}
 
	\includegraphics[angle=0,trim={0.0cm 0.0cm 2.0cm 2.1cm},clip=true,width=0.45\linewidth,keepaspectratio, page=2]{./Mdot_bin0.pdf}
 	\includegraphics[angle=0,trim={0.0cm 0.0cm 0.0cm 2.1cm},clip=true,width=0.48\linewidth,keepaspectratio, page=3]{./Mdot_bin0.pdf}
  
\caption{\normalfont{\textbf{Top-Left:} Radial distribution of the volume filling factor (FF) derived from the \ha\ and [\sii] emission lines. The best-fit power-law is shown as the blue dashed line. \textbf{Top-Right:} The black line represents the radial distribution of mass outflow rates (\Mdot) derived in Section \ref{sec:rate}. The blue line represents the de-projected outflow velocity (\Vout) from \ha\ \citep{Shopbell98}. \textbf{Bottom:} Kinetic energy and momentum outflow rates derived in Section \ref{sec:rate}.} }
\label{fig:OutflowRate}
\end{figure*}

\subsection{Radial Distribution of Outflow Rates}
\label{sec:rate}
To estimate the amounts of mass/energy/momentum carried by the warm ionized outflows per unit time, we can calculate various outflow rates. From the definition, we have:

\begin{equation}\label{eq:MPEdot}
    \begin{aligned}
    \dot{M}_\text{out} (r) &= \frac{dM}{dt} = \frac{dM}{dr} \cdot \frac{dr}{dt} = \frac{dM}{dr} V_\text{out} (r)\\
    \dot{E}_\text{out} (r) &= \frac{1}{2}\times \dot{M}_\text{out}(r) \times V^{2}_\text{out}(r)\\
    \dot{P}_\text{out} (r) &= \dot{M}_\text{out}(r) \times V_\text{out}(r)
    \end{aligned}
\end{equation}
where \Mdot, \Edot, and \Pdot\ are the mass, energy, and momentum rates of outflows for a certain radial bin, respectively; $M$ and \Vout$(r)$ are the mass and velocity of the outflows at this bin, respectively. 



For \Vout, we first adopt the observed line-of-sight (LOS) outflow velocities ($V_\text{obs}$) from \cite{Shopbell98}, where the measurements are based on detailed maps of \ha\ emission lines in the southern side of the galaxy\footnote{The northern side of M 82 is receding from us and is more dusty due to the obscuration by the disk itself. Therefore, measurements of \Vout\ and other related parameters are more reliable in southern outflow of M 82 \citep[e.g.,][]{Contursi13}, which is our focus in this paper.}. We note the deprojected outflow velocity depends on the orientation of the galaxy and outflows, i.e., \Vout\ = DF $\times$ $V_\text{obs}$, where \Vout\ is the deprojected outflow velocity in the rest-frame of M 82 and DF is the deprojection factor. Based on the radial velocity profiles of the double-peaked \ha\ emission mapped over the entire outflow, \cite{Shopbell98} found the best model to fit the data has the geometry as a pair of cones arranged as funnels with DF $\sim$ 2 (see their Section 4.3.4). We adopt their DF hereafter in this paper.




Then we can calculate the volume of a partial cone for a certain radial bin with $r_{min}$ and $r_{max}$ as the lower and upper radii by:

\begin{equation}\label{eq:dVdR}
    \begin{aligned}
    dVol &= \frac{1}{3}\pi dr \left ( \left ( \frac{W_\text{min}}{2} \right )^{2}  + \frac{W_\text{min}}{2}\frac{W_\text{max}}{2} + \left ( \frac{W_\text{max}}{2} \right )^{2} \right )\\
    &= \frac{3}{16}\pi dr \times (r^{2}_{min} + r_{min}r_{max} + r^{2}_{max} )
    \end{aligned}
\end{equation}

In step 2 above, we have adopted $W = 1.5 r$ as we measured from \ha\ outflows in Section \ref{sec:width}. Then we can calculate $dM/dr$ for each radial bin:
\begin{equation}\label{eq:dMdR}
    \begin{aligned}
    \frac{dM}{dr} = \frac{dVol}{dr} \times \text{FF} n_\text{e} \mu_{e}
    \end{aligned}
\end{equation}
where $\mu_{e}$ is the average atomic mass per electron, and FF is the volume filling factor of \ha\ emitting outflows for this bin.

    
Given \ne\ has been derived for each radial bin in Section \ref{sec:density}, the only unknown here is FF. We calculate FF from the \ha\ profiles as: 

\begin{equation}\label{eq:FF}
    \begin{aligned}
    \text{FF} = \frac{\text{F}(H\alpha)}{n_\text{e}n_\text{p}W \times \alpha_{H\alpha}E_{H\alpha}}
    \end{aligned}
\end{equation}
where F(\ha) is the average surface-brightness of \ha\ for a certain radial bin, \np\ is the proton number density and we assume \np\ = \ne/1.1 for fully ionized gas, $\alpha_{H\alpha}$ = 7.88$\times$ 10$^{-14}$ cm$^{3}$ s$^{-1}$ is the recombination coefficient of \ha\ assuming T = 10,000 K \citep{Draine11},  and $E_{H\alpha} = 3.0 \times 10^{-12}$ ergs is the \ha\ photon energy. For F(\ha), we measure it from the surface brightness of \ha\ given Subaru/FOCAS spectra \citep{Yoshida19} and correct it by the dust extinction measured from Balmer decrements \citep{Heckman90}. As noted above, due to the higher/uncertain dust extinction on the north-west side, we only compute a dust-corrected \ha\ surface brightness for the south-east side of the outflows. We will finally multiply the resulting outflow rates by a factor of two to represent the total amounts for M 82 (see below).

The resulting radial distribution of FF is shown in the first panel of Figure \ref{fig:OutflowRate}. We find that FF steadily decreases from 0.3\% at $r$ = 0.5 kpc to 0.005\% at 2.2 kpc, which shows that the warm ionized clouds are extremely clumpy, and occupy less volume as they travel further out of the galaxy. The best-fit power-law is given by FF$(r) = 10^{-3} \times (\frac{r}{628 pc})^{-1.8}$ and is shown as the blue dashed line in Figure \ref{fig:OutflowRate}.

Overall, combining Equations (\ref{eq:MPEdot}) -- (\ref{eq:FF}), we can get the radial distributions of \Mdot, \Edot, and \Pdot\ for the warm-ionized outflows in M 82 (Figure \ref{fig:OutflowRate}). We find the outflow rates drop quite slowly and stay almost unchanged within the error bars from 0.8 to 2.2 kpc. This suggests that the galactic wind in M 82 can indeed carry the mass, energy, and momentum from the central regions out to a few kpc with minimal losses. 




\section{Discussion}
\label{sec:discuss}

\subsection{Comparisons of Spatially-resolved Multi-phase Outflow Rates in M 82}
\label{sec:CompOutflowRates}
 Detailed spatially-resolved studies of the M 82 wind exist for various outflowing phases. Here we will compare our measured outflow rates as a function of radius from \ha\ and [\sii] to the published values for these other phases. Specifically, we will compare to maps of the outflow rates of cold atomic gas \citep{Martini18} and cold molecular gas \citep{Leroy15}. Estimating these rates requires a measurement of the intrinsic outflow velocity. Given the nearly edge-on orientation of M 82, there is a significant correction needed to convert observed line-of-sight outflow velocities to intrinsic values. To be able to compare the outflow rates in the different phases in a consistent way, we use the deprojection factor (DF) adopted for the warm ionized gas to measure outflow rates in the previous studies, i.e., DF = 2 \citep[][see Section \ref{sec:rate} above]{Shopbell98}. We have updated their measured outflow rates accordingly. We list the results in Table \ref{tab:stat} and briefly discuss these values as follows.

Spatially-resolved maps of the cold neutral outflowing gas in M 82 via spatially resolved \hi\ 21 cm emission were discussed by \citet{Martini18}. Adopting DF $=$ 2, the outflow velocities decline with radius from about 200 km s$^{-1}$ at 1 kpc to only 50 km s$^{-1}$ at 10 kpc. There is a corresponding steep decline in the mass outflow rates. Over the radial range we probe for the warm ionized gas (out to 2.2 kpc), the mass outflow rates in HI are similar to those for the warm ionized gas, but the outflow rates of momentum and (especially) kinetic energy are significantly smaller (Table \ref{tab:stat}).  

Maps of outflowing cold molecular gas in M 82 based on observations of CO emission lines were described by \cite{Leroy15}. For DF $=$ 2, the inferred \Vout\ is about 150 km s$^{-1}$, which is significantly smaller than the values in the warm ionized phase, but similar to the values for the atomic phase.  (Table \ref{tab:stat}). The implied mass outflow rates decline from about 10 M$_{\odot}$ yr$^{-1}$ at a radius of 1 kpc to values over an order-of-magnitude smaller by a radius of 3 kpc. Comparing these to the the outflow rates for the warm ionized gas over the the radial range we probe, the mass outflow rates probed by CO are two times larger, the momentum outflow rates are about three times smaller, and the kinetic energy outflow rates are about ten times smaller.

Taken together, the relatively small outflow velocities and the steep decline in outflow rates with radius are consistent with a picture in which the atomic and molecular gas traces a fountain flow that launches gas out to a few kpc \citep{Leroy15}.
We also note that the combined outflow rates of the three phases amount to about 50\% of the momentum injected by the M 82 starburst, and only about 16\% of the injected kinetic energy. These results are consistent with the finding that the very hot gas (3 -- 8 $\times 10^{7}$ K) in M 82 that is feeding the fast wind fluid carries the rest of the momentum flux and nearly all the kinetic energy flux \citep{Strickland09}.


\begin{table*}
	\centering
	\caption{Comparisons of Spatially Resolved Outflow Rates from Different Phases in M 82$^{(*)}$}
	\label{tab:stat}
	\begin{tabular}{llllllll} 
		\hline
		\hline
		Phases & Tracer & Log(\Vout)$^{(a)}$ & Log(\Mdot) & Log(\Pdot)  & Log(\Edot) & Radii$^{(b)}$ &Reference\\
		   & & (km/s) & (\Msun/yr) & (dynes) & (ergs/s) & (kpc) &\\
		\hline
		\hline
  
            Warm ionized   & \ha        & 2.8 & 0.8 -- 0.0 & 34.3 -- 33.9 & 41.7 -- 41.4 & 0.7 -- 2.2 & This paper\\

            Cold atomic    & \hi\ 21cm  & 2.3 & 0.5 -- 0.1 & 33.6 -- 33.2 & 40.6 -- 40.2 & 1.0 -- 2.2 & \cite{Martini18}\\

            Cold molecular & CO         & 2.2 & 1.0 -- 0.3 & 34.0 -- 33.3 & 40.9 -- 40.2 & 1.0 -- 2.2 & \cite{Leroy15}\\

		\hline
		\hline
	\multicolumn{8}{l}{%
  	\begin{minipage}{15cm}%
	Note. -- \\
           \textbf{(*).}\ \ For the cold atomic and molecular gas, we assume that the deprojection factor is the same as applied to the warm ionized gas \citep[DF = 2, see][and Section \ref{sec:rate}]{Shopbell98}. \\
           \textbf{(a).}\ \ These outflow velocities are used with the mass outflow rates to compute \Pdot\ and \Edot. For \ha, we adopt the \Vout\ from \cite{Shopbell98}.\\   
           \textbf{(b).}\ \ The values for all parameters cover the corresponding range in radii shown in this column.\\  
           \textbf{(c).}\ \ The relevant values for the M 82 starburst are Log(SFR) = 0.9 (\Msun/yr), Log(\PdotStar) = 34.6 (dynes), and Log(\EdotStar) = 42.5 (ergs/s).\\  
  	\end{minipage}%
	}\\
	\end{tabular}
	\\ [0mm]
	
\end{table*}

\begin{figure*}
\center
	\includegraphics[page = 1, angle=0,trim={0.2cm 0.2cm 0.20cm 0.0cm},clip=true,width=0.5\linewidth,keepaspectratio]{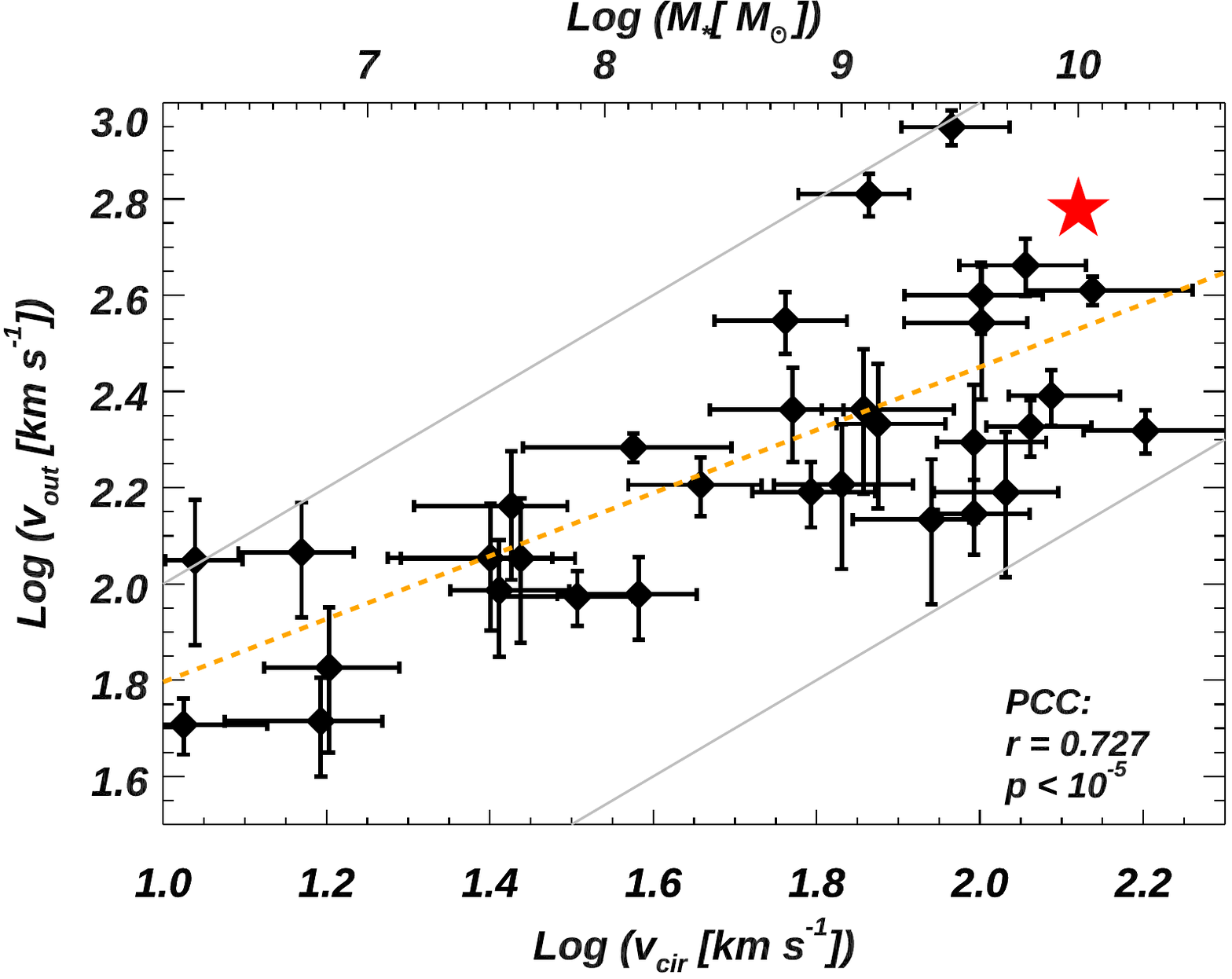}
	\includegraphics[page = 1, angle=0,trim={0.2cm 0.2cm 0.20cm 1.0cm},clip=true,width=0.5\linewidth,keepaspectratio]{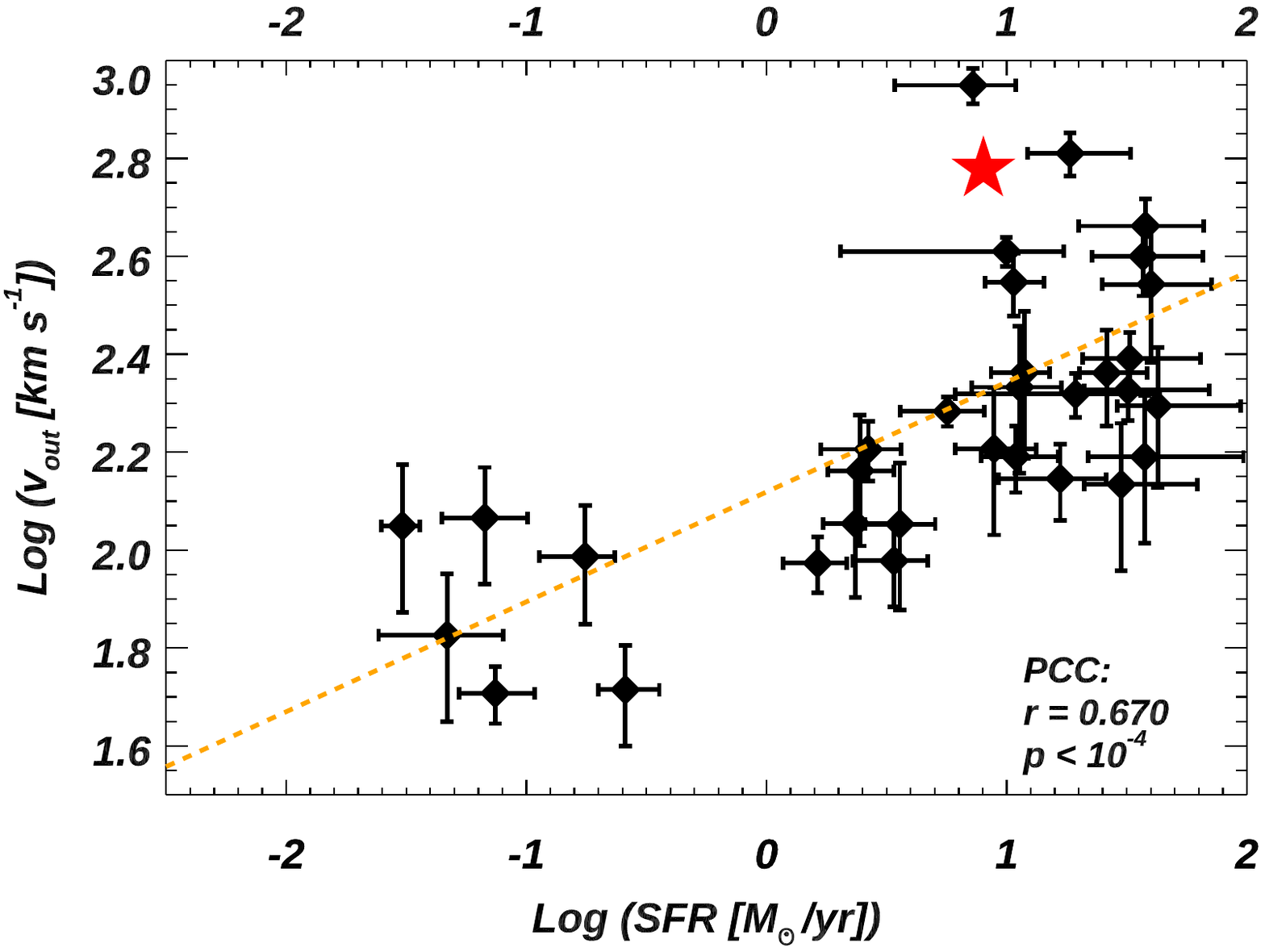}

\caption{\normalfont{The log of the outflow velocity (\Vout) versus circular velocity (\textbf{Left}) and star formation rate (\textbf{Right}). Galaxies from the CLASSY sample is shown in black, while measurements for in M 82 in this paper are shown as the red star. Orange dashed lines represent the best-fit linear correlation presented in \cite{Xu22a}. In general, the galactic winds in M 82 match the scaling relationships reported for low-redshift SF galaxies in CLASSY.} }
\label{fig:OutflowVoutComp}
\end{figure*}

\subsection{Comparisons of the Outflow in M 82 to those in the CLASSY Sample}
\label{sec:CompareCLASSY}

The systematic properties of warm ionized outflows have been widely studied via interstellar absorption lines \citep[e.g.,][]{Martin05, Rupke05, Chisholm15, Heckman15, Heckman16, Xu22a}. However, how the galactic outflow in M 82 as traced in emission compares to these estimates for a large sample of starburst galaxies is still an open question. Here we compare the ionized outflow properties in M 82 with the outflows observed in the COS Legacy Archive Spectroscopy SurveY (CLASSY) atlas \citep{Berg22, James22}. CLASSY includes 45 low-redshift starburst galaxies (z = 0.002 -- 0.182), which occupy a wide range of important galaxy properties, including stellar mass, SFR, and metallicity. Their outflow features and correlations with the galaxy properties have been analyzed in a homogeneous way and are reported in \cite{Xu22a} and \cite{Xu23}. These results are based on spatially unresolved FUV spectra from the Hubble Space Telescope (HST)/Cosmic Origins Spectrograph (COS).

\begin{figure*}
\center
	\includegraphics[page = 1, angle=0,trim={0.1cm 0.2cm 0.1cm 0.0cm},clip=true,width=0.5\linewidth,keepaspectratio]{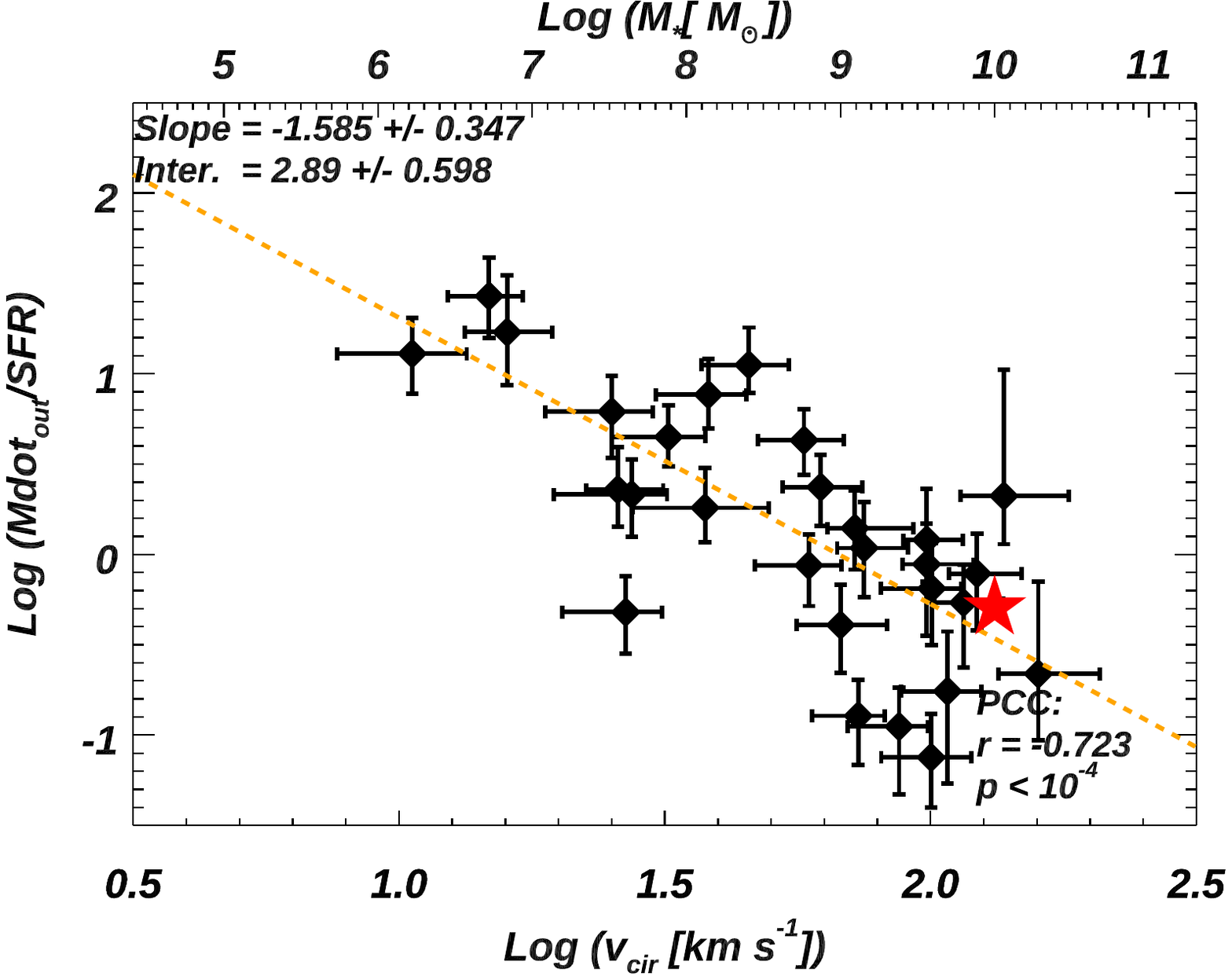}
	\includegraphics[page = 1, angle=0,trim={0.1cm 0.2cm 0.1cm 0.0cm},clip=true,width=0.5\linewidth,keepaspectratio]{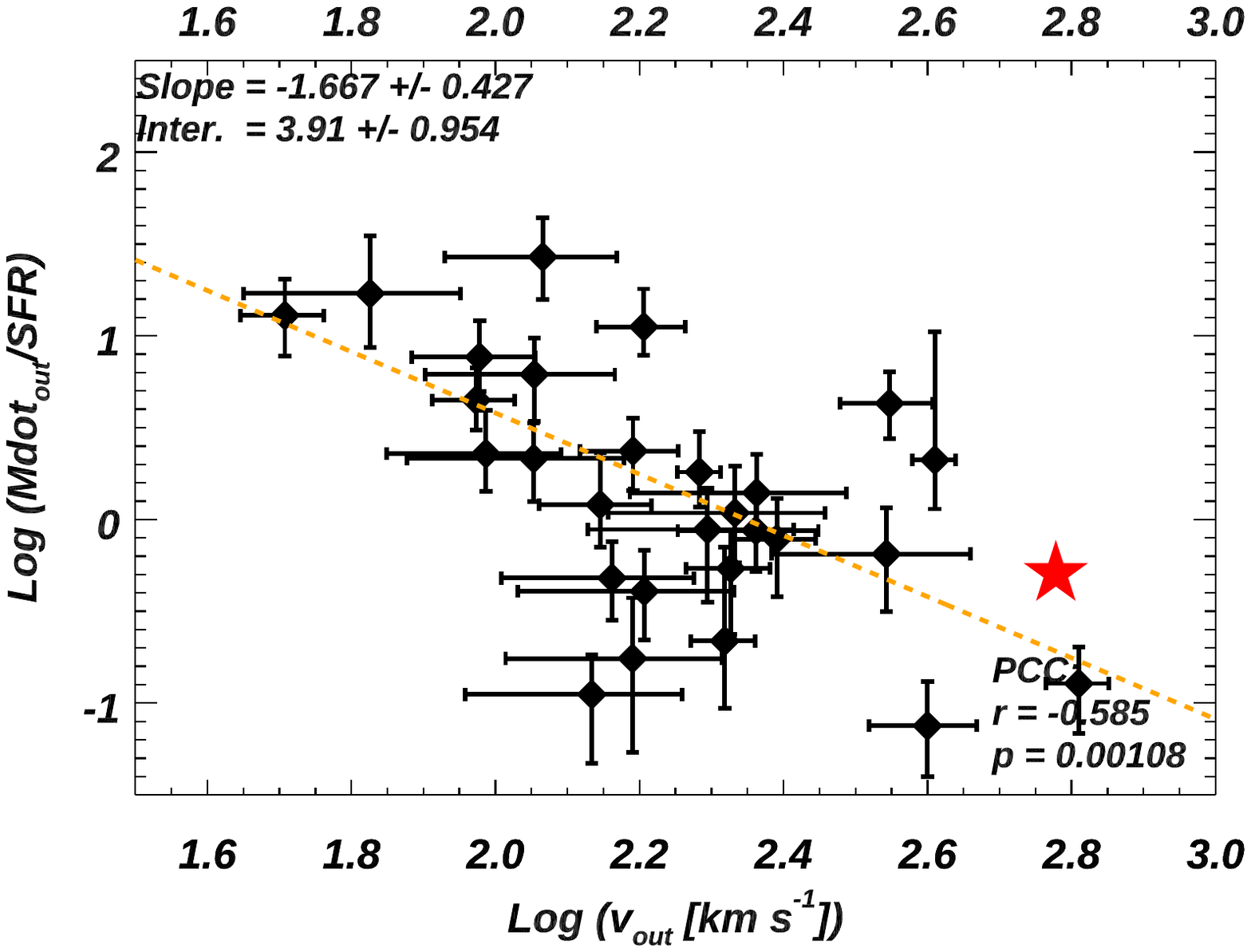}
	
	\includegraphics[page = 1, angle=0,trim={0.1cm 0.2cm 0.1cm 0.3cm},clip=true,width=0.5\linewidth,keepaspectratio]{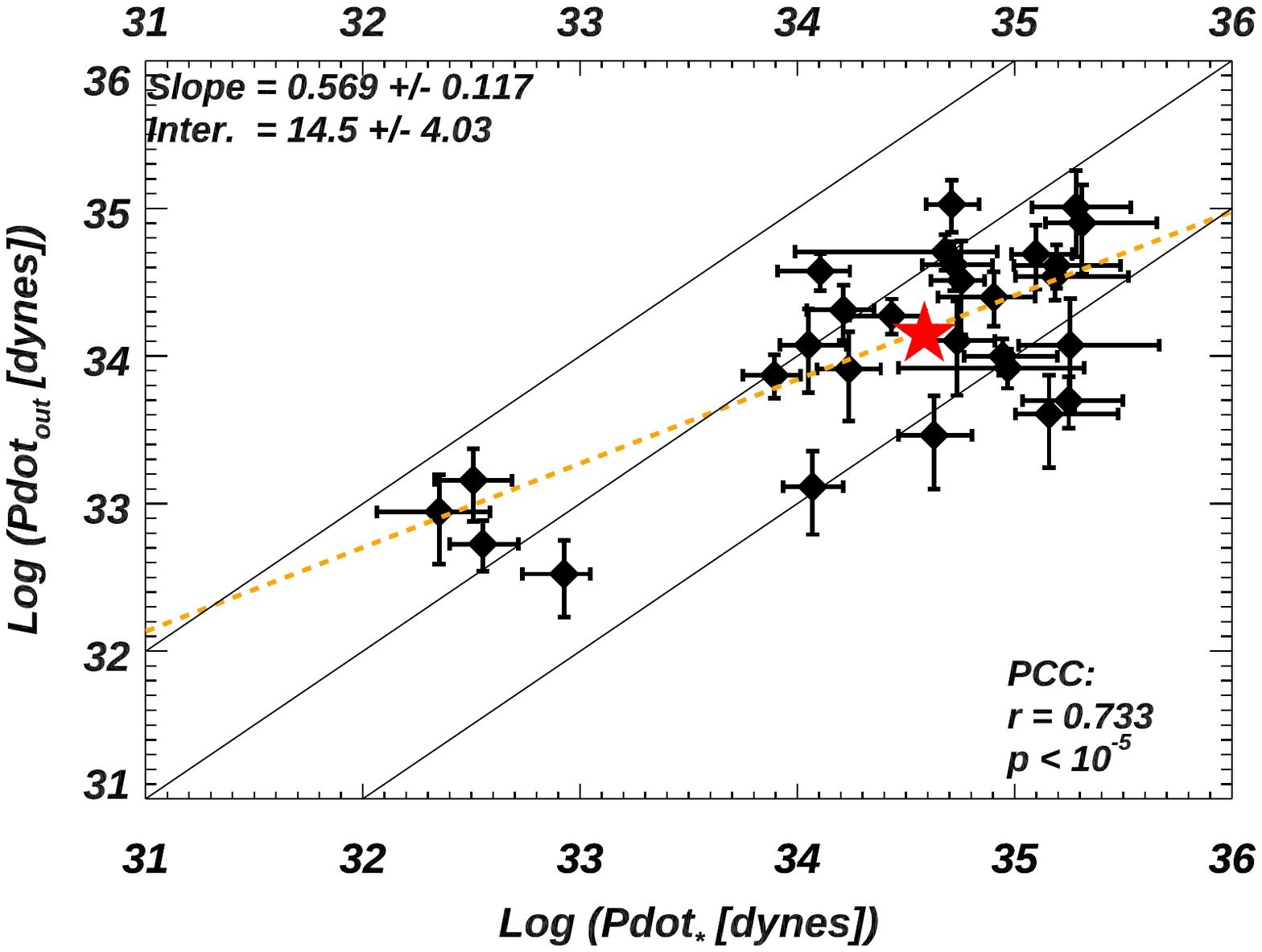}
	\includegraphics[page = 1, angle=0,trim={0.1cm 0.2cm 0.1cm 0.3cm},clip=true,width=0.5\linewidth,keepaspectratio]{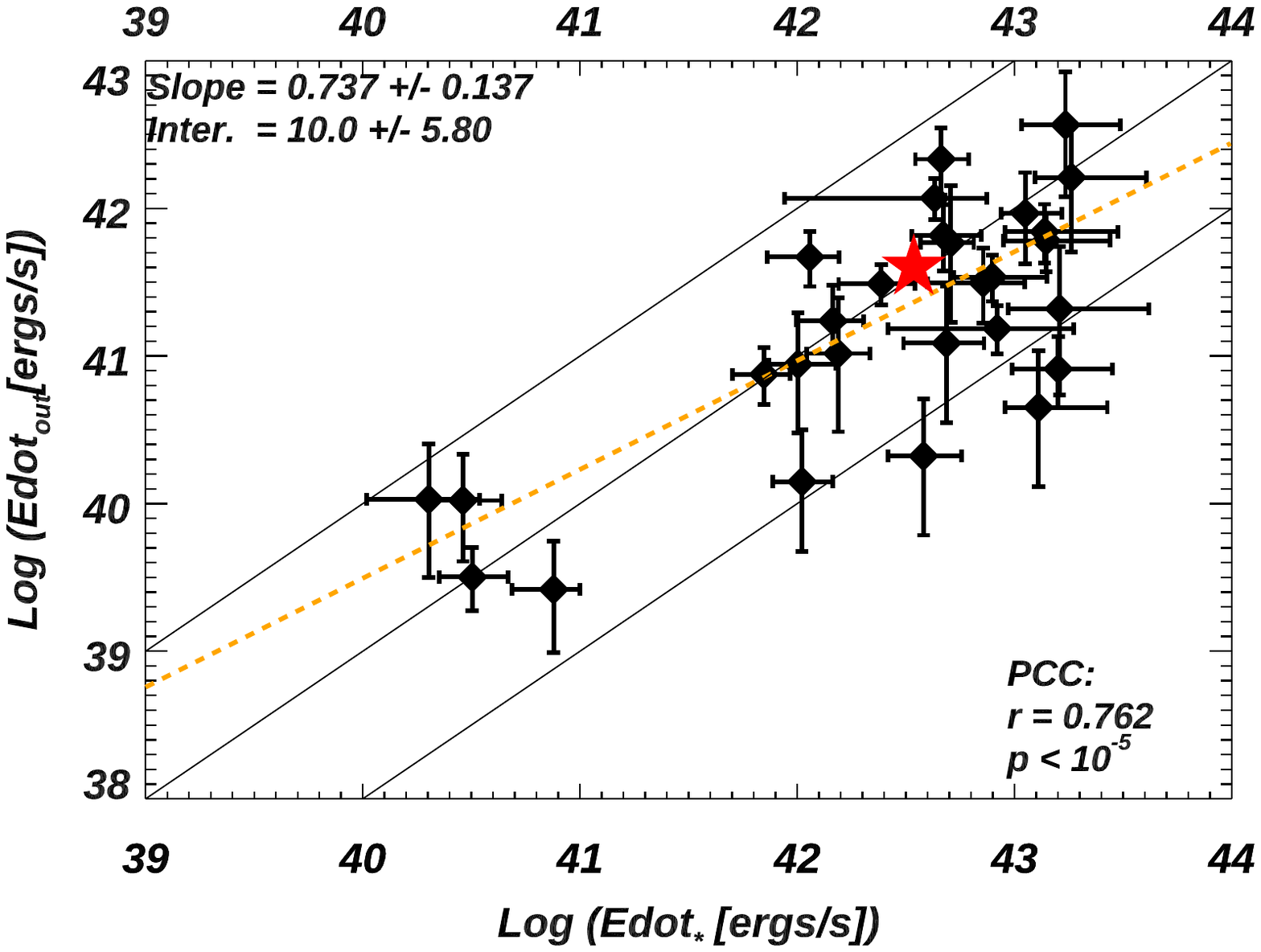}

\caption{\normalfont{\textbf{Top:} Correlations related to the mass loading factor (\Mdot/SFR). Labels and captions are the same as Figure \ref{fig:OutflowVoutComp}. \textbf{Bottom:} Correlations of the momentum (energy) outflow rates versus momentum (energy) suppied by the starburst. See disucssion in Section \ref{sec:CompareCLASSY}.} }
\label{fig:MdotCorr}
\end{figure*}

In Figure \ref{fig:OutflowVoutComp}, we show the outflow velocity versus two important galaxy properties, i.e., stellar mass (left) and SFR (right). The measurements of warm ionized outflows (Section \ref{sec:analysis}) in M 82 are shown as red stars, where we take 600 km s$^{-1}$ that is valid for $r$ $>$ 0.7 kpc (Figure \ref{fig:OutflowRate}). In general, we find the ionized outflow in M 82 studied in emission-line matches the scaling relationships derived from the CLASSY sample based on absorption-line data. 

Similarly, in Figure \ref{fig:MdotCorr}, we compare the integrated (total) mass, momentum, and energy outflow rates with the mass, momentum, and energy input (i.e., SFR, \PdotStar, \EdotStar) provided by the starburst regions, respectively \citep{Xu22a}. For M 82, we show the mean values for radii between 0.5 and 2.2 kpc. We find these outflow rates in M 82 well match the trends reported for the CLASSY sample. In the top panels, we also show the mass-loading factor (\Mdot/SFR) versus \Vout\ and \Vcir, where M 82 is located at the bottom-right corner. Again, the location of M 82 in these plots is consistent with the CLASSY sample.


Furthermore, the median value of FF for CLASSY galaxies is 4 $\times 10^{-3}$. This is similar to the one for M 82 ($\sim 10^{-3}$ to 10$^{-4}$ , Figure \ref{fig:OutflowRate}). Given the best-fit \ne(r) and FF(r) in Figures \ref{fig:OutflowDensity} and \ref{fig:OutflowRate}, we find FF$\times n \propto r^{-3}$. Then we can also derive the LOS integrated outflow column density from \ha+[\sii] observations as: 

\begin{equation}\label{eq:NHLOS}
    \begin{aligned}
    N_\text{H,LOS} (r) &= \int_{500 pc}^{\infty}{\text{FF}\times n \times dr} = 3.2 \times 10^{20} cm^{-2}
    \end{aligned}
\end{equation}

This value is quite close to the median $N_\text{H,LOS}$ = 4.9 $\times 10^{20}$ cm$^{-2}$ measured in the CLASSY sample.

Overall, we find that the warm ionized outflows probed by \ha\ emission lines in M 82 follow the same scaling relationships and have similar LOS column densities as those reported in the CLASSY sample \citep{Xu22a, Xu23} for warm ionized outflows studied in absorption. This suggests that the outflow properties in M 82 are similar to those in other low-redshift starburst galaxies, and that the ionized gas seen in emission and absorption is likely to trace similar material. We summarize all these comparisons in Table \ref{tab:GlobalOutflow}\footnote{We refer readers to Table 3 in \cite{Xu22a} for more comparisons between CLASSY and other low-redshift starburst galaxy samples.}.

\begin{table}
	\centering
	\caption{Comparisons of Outflows in M 82 and CLASSY$^{(*)}$}
	\label{tab:GlobalOutflow}
	\begin{tabular}{llrr} 
		\hline
		\hline
		Parameters & Unit & M 82 & CLASSY   \\
		 (1) & (2) & (3) & (4) \\
		\hline
		\hline

		Log(\Mdot/SFR)    & (1) & --0.30 & --0.60 \\
		Log(\Vout/\Vcir)  & (1) & 0.60   & 0.55 \\
            Log(\Pdot)        & (dynes) & 34.1   & 34.2 \\
            Log(\Edot)        & (ergs s$^{-1}$) & 41.6   & 41.3 \\
            Log(N$_\text{H,LOS}$)$^{(a)}$     & (cm$^{-2}$) & 20.5   & 20.7 \\
            \hline
		Log(\ne)          & (cm$^{-3}$) & 1.7 to 2.3   & 1.5 \\
		Log(FF)           & (1) & --2.6 to --4.1   & --2.5 \\
            Log(\Rcloud)$^{(b,c)}$      & (pc) & --1.2 to --0.0   & 0.7 \\
            Log(\NHcloud)$^{(b,c)}$     & (cm$^{-2}$) & 19.1 to 20.7   & 20.8 \\


		\hline
		\hline
	\multicolumn{4}{l}{%
  	\begin{minipage}{8cm}%
	Note. --\\
	    \textbf{(*).}\ \ For M 82 values related to outflow rates (first four rows), we show the mean values between radii of 0.5 and 2.2 kpc. For M 82's outflow cloud properties (latter four rows), we present the values correspond to the range in radii of 0.5 -- 2.2 kpc. For CLASSY sample, we list their published median values \citep{Xu22a, Xu23}. See discussion in Section \ref{sec:CompareCLASSY}.\\
        \textbf{(a).}\ \ The line-of-sight integrated hydrogen column density (see Equation \ref{eq:NHLOS}).\\
	\textbf{(b).}\ \ The cloud properties for M 82 are derived in Section \ref{sec:structure}.\\
  	\textbf{(c).}\ \ The cloud radii and column densities for M 82 assume that the cloud covering factor has the median value derived for the CLASSY sample (see equation 6).\\
   \end{minipage}%
	}\\
	\end{tabular}
	\\ [0mm]
	
\end{table}

\subsection{Constraints on Outflow Cloud Parameters}
\label{sec:structure}
Outflows are in the form of separate clouds, whose sizes (\Rcloud), masses (\Mcloud), and column densities (\NHcloud) are key parameters to determine whether they can survive long enough to be accelerated by the hot wind \citep[e.g.,][]{Gronke20, Li20, Sparre20, Kanjilal21, Abruzzo22, Fielding22}. Thus, these parameters are critical to decipher if outflows can still be significant at large scales within a galaxy. Nonetheless, observational constraints of them are rare in the literature \citep[except in][]{Xu23}. 

As shown in \cite{Xu23}, one can estimate \Rcloud\ and \NHcloud\ as follows [see their Section 4.4]:

\begin{equation}
\label{eq:R}
    \begin{aligned}
    R_\text{cl} (r) &= \frac{3}{4} \frac{\text{FF}}{\text{CF}_{sh}} L(r)\\
    N_\text{H,cl} (r) &= R_\text{cl} (r) \times n_\text{H, cl} (r)
    \end{aligned}   
\end{equation}
where \text{L}(r) is the line-of-sight (LOS) path-length through the outflow, CF$_{sh}$ = CF/$\beta_{sh}$ is the outflow LOS covering factor after accounting for shadowing effects. This is because the projected areas by different outflow clouds within the LOS can overlap so that the total covered area drops by a factor of $\beta_{sh}$.

\begin{figure*}
\center
	\includegraphics[angle=0,trim={0.0cm 0.0cm 0.0cm 0.5cm},clip=true,width=0.45\linewidth,keepaspectratio,page=2]{./NH_RCloud_bin0.pdf}
	\includegraphics[angle=0,trim={0.0cm 0.0cm 0.0cm 0.5cm},clip=true,width=0.45\linewidth,keepaspectratio,page=1]{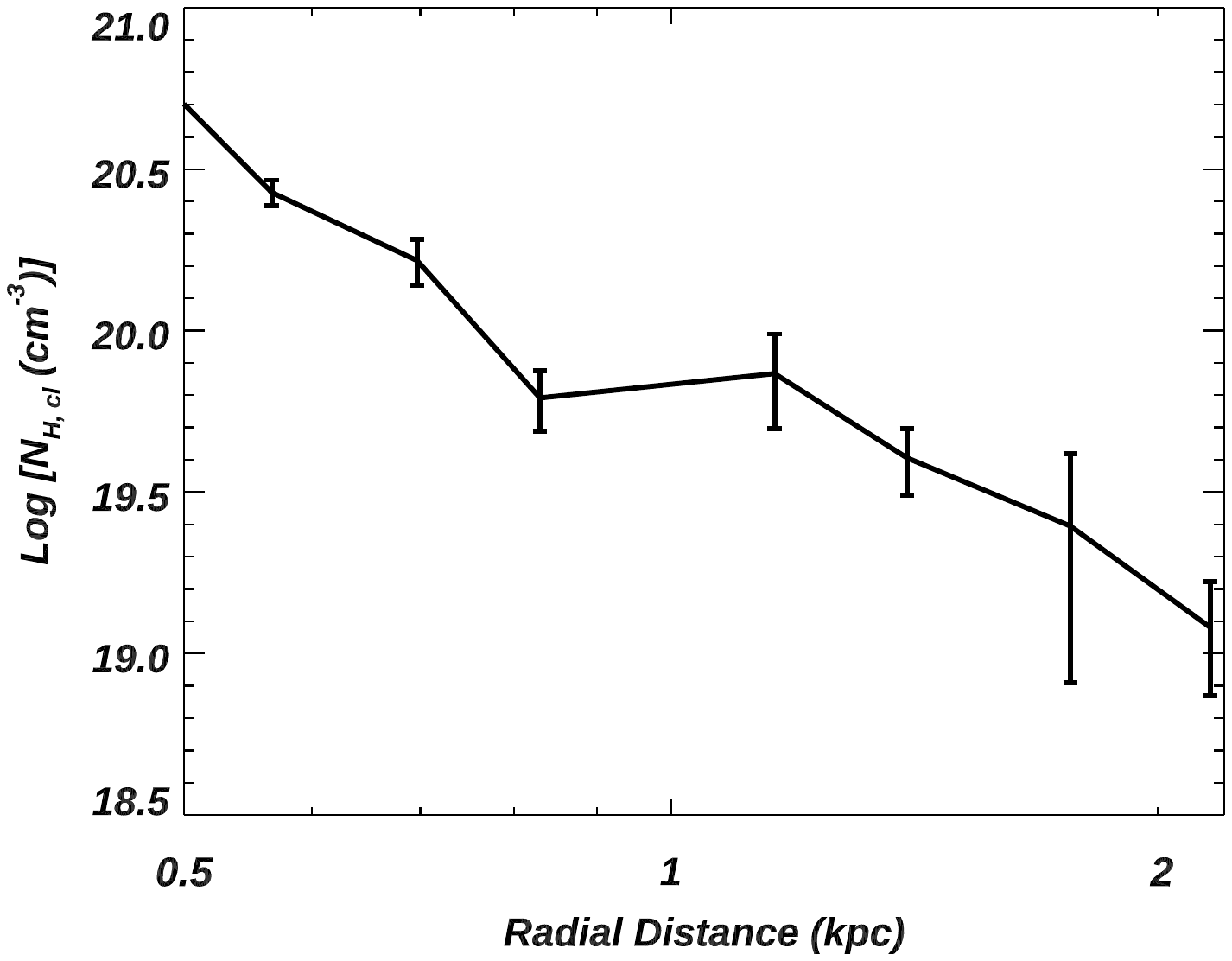}\\
\caption{\normalfont{\textbf{Left:} Radial distribution of the outflow cloud sizes. 
\textbf{Right:} Radial distribution of the cloud hydrogen column density (\NHcloud) for the outflows. See Section \ref{sec:structure} for details.} }
\label{fig:OutflowClouds}
\end{figure*}

In M82, we do not have direct estimates of CF$_{sh}$. Given the similarities of outflows seen in M82 and CLASSY galaxies (see Section \ref{sec:CompareCLASSY}), we adopt the median value of CF$_{sh}$ from CLASSY sample as a rough estimate ($\sim$ 1.7). Overall, from radius of 0.5 to 2 kpc, we get \Rcloud\ = 0.9 -- 0.07 pc and \NHcloud\ = $10^{20.7}$ --  $10^{19.1}$cm$^{-2}$. We also show the distributions for them in Figure \ref{fig:OutflowClouds}.

We can compare the above derived warm-ionized outflow cloud properties to the results in \cite{Krieger21}, which attempted to constrain the molecular cloud properties in M 82 based on CO(1--0) observations. They find the molecular clouds are much larger (50$\pm$10 pc) and slower moving than the ionized clouds. Comparing with galaxies in CLASSY sample, our measured \Rcloud\ and \NHcloud\ values are also 0.2 -- 2 dex smaller (see the last two rows in Table \ref{tab:GlobalOutflow}).

\subsection{Comparisons with Theory}
\label{sec:CompTheory}
\subsubsection{Description of Models and Simulations}
\label{sec:desModel}
The results we have presented can be compared to recent models and simulations of outflows that are designed to capture the physical processes occurring in the multi-phase galactic winds. Here we focus on two recent investigations of multi-phase galactic winds, namely the semi-analytic models by \citet{Fielding22} and the high-resolution numerical simulations by \cite{Schneider20} \footnote{There are other recent numerical simulations of outflows \citep{Kim20,Steinwandel22,Rey23}, however these simulations are very poor matches to M 82 in terms of the galaxy mass, SFR, and SFR/A.} In both cases, there is an important distinction between the tenuous and high-velocity "wind fluid" that is created by the thermalized ejecta (winds and supernovae) of massive stars, and the denser, slower-moving ambient gas with which it interacts ("outflowing clouds"). The reported warm ionized outflow through \ha\ in this paper represents only the latter phase.  \citet{Fielding22} and \cite{Schneider20} represent significant improvements on previous models and simulations. The \citet{Fielding22} model is the first to incorporate physically-based mechanisms for the exchange of mass, momentum, and energy between the clouds and the wind. The \citet{Schneider20} simulations combine significantly-improved spatial resolution that better captures the underlying physics, and more realistic treatments of how mass and energy are injected by the massive stars.

\begin{figure*}
\center
	\includegraphics[angle=0,trim={0.0cm 0.0cm 0.0cm 0.0cm},clip=true,width=1.0\linewidth,keepaspectratio,page=1]{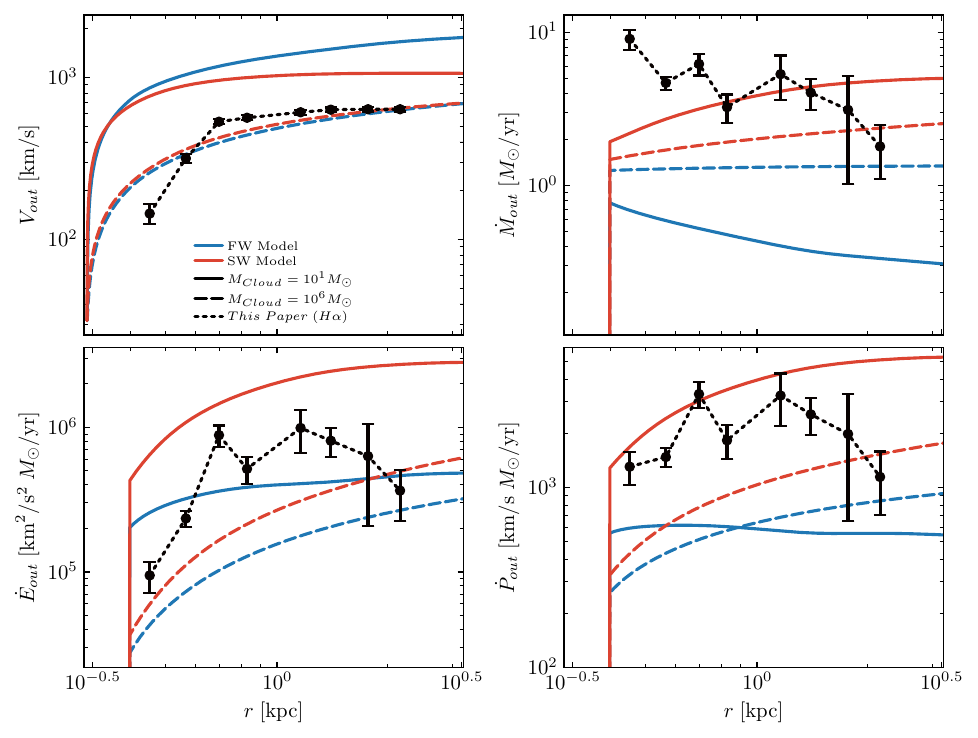}
\caption{\normalfont{Comparisons of the semi-analytic models by \cite{Fielding22} with the derived outflow parameters in this paper. We have rerun their models to match the input parameters of M 82 (i.e., SFR = 8 M$_{\odot}$ yr$^{-1}$, starburst radius = 300 pc, and bi-polar outflows with opening angle of 73.7$^{\circ}$). The blue and red lines represent the models with initial hot phase mass loading factor as 0.1 and 0.5, respectively. The former produces faster and less dense hot wind fluid (hereafter, FW model), while the latter generates slower and denser wind fluid (hereafter, SW model). The solid and dashed lines represent different outflow cloud mass into the models, i.e., 10$^{1}$ and 10$^{6}$ \Msun, respectively. In the four panels, we show the model predicted outflow cloud velocity, mass, energy, and momentum outflow rate, respectively. In each panel, we overlay our observed outflow properties from \ha\ as black dotted lines with errorbars. We discuss these comparisons in detail in Section \ref{sec:CompTheory}.}}
\label{fig:CompModelFB}
\end{figure*}

In the \citet{Fielding22} semi-analytic model, the wind fluid created by stellar ejecta in the starburst interacts with a population of pre-existing clouds in which there can be a two-way exchange of mass, momentum, and energy. To match the exact conditions of M 82, we have rerun their models with SFR = 8 M$_{\odot}$ yr$^{-1}$, starburst radius = 300 pc, and bi-polar outflows with opening angle of 73.7$^{\circ}$ (as derived in Section \ref{sec:width}). We present the results of four settings in Figures \ref{fig:CompModelFB} and \ref{fig:CompModelFB2} (colored lines) and compare them with our observed outflow properties from \ha\ (black dotted lines). We discuss the comparisons in details over the next few subsections.



The simulation in \citet{Schneider20} starts with a wind-fluid created in a super-star-clusters located inside a starburst with a radius of 1000 pc and SFR = 20 M$_{\odot}$ yr$^{-1}$. The starburst is embedded in a gaseous disk with a radial exponential scale-length of 1.6 kpc and a gas mass of 2.5 $\times 10^9$ M$_{\odot}$. The spatial resolution is 5 pc. There is a temperature floor at 10$^4$ K, so that the warm phase will be more homogeneous in temperature in the simulation than in reality. In the simulations the collimating effects of the disk lead to a bi-polar outflow that resembles the one in M 82.

\subsubsection{Outflow Properties}
\label{sec:CompOutflowProperties}
We begin by comparing the radial profiles of the outflow velocity (\Vout) and outflow rates of the warm ionized gas in M~82 to the predictions of \citet{Fielding22} and \citet{Schneider18}. The former is shown in Figure \ref{fig:CompModelFB} and the latter is summarized in Tables \ref{tab:CompVout}.

For models by \citet{Fielding22}, blue and red lines represent models that produce faster, less denser winds and slower, denser winds, respectively (hereafter, FW and SW models). The dashed and solid lines represent different initial outflow cloud masses \footnote{The outflow clouds in the model by \citet{Fielding22} are added gradually into the model at $r$ = 300 -- 400 pc}. Our measured values from \ha\ in M 82 are shown as the black dotted lines. We find the ones with \Mcloud\ = 10$^{1}$ \Msun\ predict larger velocities than are observed, and the agreement is better for the most massive clouds (\Mcloud\ $ =10^6$ M$_{\odot}$).  
The model with a slower and denser wind fluid (red curves) predicts lower velocities (more consistent with M 82). For \Mdot, the FW model with lower mass clouds (blue solid line in the second panel) produce \Mdot\ values that do not match the data for M82: the rates are too small and decline too quickly with radius. Their SW model (red lines) shows a better match to our measurements. For \Edot, the FW model with lower \Mcloud\ and SW model with higher \Mcloud\ match better with the data. For \Pdot, the SW model with initial \Mcloud\ between our two cases should match the data better.

On the other hand, we find that the outflow velocities are under-predicted in the \citet{Schneider20} simulation (Tables \ref{tab:CompVout}). The simulation also shows a steady increase in outflow velocity with increasing radius, which is barely seen in M 82. In the columns (4) and (5) of Table \ref{tab:CompVout}, the values of \Mdot\ derived for the warm ionized gas (normalized for the SFR) in the \citet{Schneider20} simulation are significantly smaller than the values measured for M 82.

\begin{table*}
	\centering
	\caption{Comparisons of Outflows Properties from M 82 and \cite{Schneider20}$^{(*)}$}
	\label{tab:CompVout}
	\begin{tabular}{lccccccc} 
		\hline
		\hline
		Radius & V$_\text{out, S+}$   & V$_\text{out, M 82}$ & (\Mdot/SFR)$_\text{S+}$   & (\Mdot/SFR)$_\text{M 82}$ & $P_\text{th, S+}$ & $P_\text{ram, S+}$ & $P_\text{M 82}$\\
		 (1) & (2) & (3) & (4) & (5) & (6) & (7) & (8) \\
		\hline
		\hline

		500     & \dots        &  200  & \dots  &  0.37 & \dots  & \dots &  4.2 $\times$ 10$^{6}$\\
		1000    & 100 to 300   &  600  & 0.02   &  0.25 & 1.0 $\times$ 10$^{5}$   &  4.6 $\times$ 10$^{5}$ &  1.8 $\times$ 10$^{6}$\\		
            2000    & 200 to 500   &  600  & 0.03   &  0.13 & 1.3 $\times$ 10$^{4}$   &  2.1 $\times$ 10$^{5}$ &  9.0 $\times$ 10$^{5}$\\

		\hline
		\hline
	\multicolumn{8}{l}{%
  	\begin{minipage}{14cm}%
	Note. --\\
	    \textbf{(*).}\ \ Radius is in unit of parsec, while all velocities are in units of km s$^{-1}$ and pressures are in units of Pa/K. The mass outflow rates are normalized by the SFR.\\
            \textbf{(2), (4), (6) and (7):}\ \ Model predictions from \cite{Schneider20}. For pressures, we have divided their values by 2.5 to correct for the lower SFR in M 82 compared to the model.\\
            \textbf{(5) and (8):}\ \ Results derived in this paper based on rest-optical observations of M 82 (see Figure \ref{fig:OutflowRate}). The pressures in M 82 are derived from the measured \ne\ (see Figure \ref{fig:OutflowDensity} and Section \ref{sec:CompDensityPressure}).
  	\end{minipage}%
	}\\
	\end{tabular}
	\\ [0mm]
	
\end{table*}

\subsubsection{Radial Density and Pressure Profiles}
\label{sec:CompDensityPressure}
\begin{figure*}
\center
	\includegraphics[angle=0,trim={0.0cm 0.0cm 0.0cm 5.8cm},clip=true,width=1.0\linewidth,keepaspectratio,page=1]{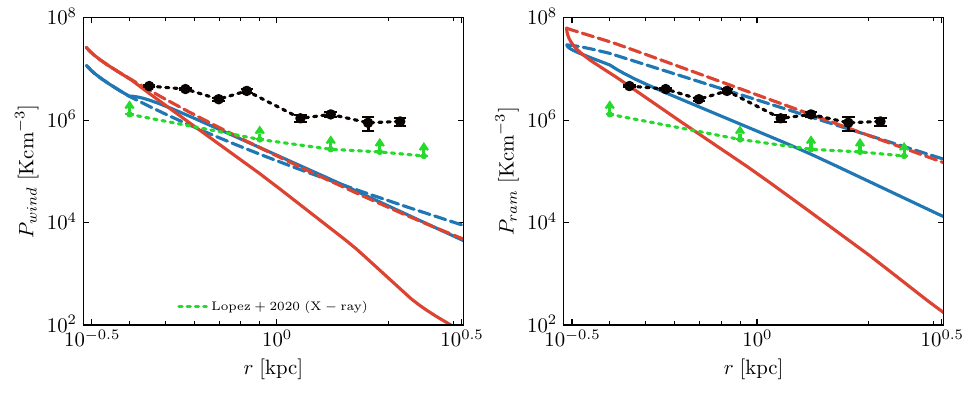}
\caption{\normalfont{Same as Figure \ref{fig:CompModelFB} but for the hot wind fluid thermal pressure (\textbf{Left}) and the ram pressure on the outflowing clouds (\textbf{Right}). In both panels, we also show the lower limits on wind pressure from the X-ray data in \cite{Lopez20} as the green lines.}}
\label{fig:CompModelFB2}
\end{figure*}

Next, we compare the measured the radial density gradient in the warm ionized outflow in M 82 (Figure \ref{fig:OutflowDensity}) to predictions from the two papers described above. \citet{Fielding22} have shown the radial profile of the gas pressure $P$ in the outflows. To convert our electron densities in M 82 to pressures ($P$), we take $P/k_\text{B} = 2 n_e T$, and assume $T \sim 10^4$ K for the warm ionized gas. This is appropriate for photoionized gas \citep{Schneider20,Xu22a}\footnote{If the emission-line gas is shock-heated, it will have a higher temperature and hence a higher inferred pressure. This will only strengthen our conclusions below.}. In the left panel of Figure \ref{fig:CompModelFB2}, it is clear that the predicted wind pressures by \cite{Fielding22} are too low in all four models. The discrepancies grow with distance, reaching about a factor of $\sim$30 to over 100 at a distance of 2 kpc. 

One possible interpretation of this would be that the densities (and hence the pressures) derived from the [\sii] doublet ratio are biased to higher-than-average values. This could occur if there is a range in density along a line-of-sight, and the [\sii] emission is weighted towards the higher density regions (since the emissivity per unit volume is proportional to $n^2$).\footnote{If the [\sii] densities are indeed biased high (by some factor $\text{B} >>1$, as required to match the thermal pressures in the models), then all the observed outflow rates in Figure \ref{fig:CompModelFB} would be boosted by $\text{B} >> 1$, and become unphysically large.} To test this, we can compare the pressures derived from [\sii] for the warm phase to those derived independently for the hot X-ray-emitting phase. This is reasonable, since there is a close morphological correspondence between the optical and soft X-ray emission in the M 82 outflow \citep{Heckman17a}. Therefore, in the same panel, we also overlay the radial wind pressure profile derived from X-ray measurements of M 82 in \cite{Lopez20} (green line). Since they assume X-ray volume filling factors ($\text{FF}_\text{X}$) of unity and since $P_\text{X} \propto \text{FF}_\text{X}^{-1/2}$, their measurements are strict lower limits, corresponding to minimum pressures $\sim$ 25\% as large as our estimates. We emphasize that the complex filamentary structure seen in the soft X-ray emission is inconsistent with unit filling factor\footnote{The pressures derived from the [\sii] ratios and the X-ray data would agree for $\text{FF}_\text{X} \sim 0.06$.}. Even if $\text{FF}_\text{X} = 1$, all the models significantly underpredict $P_\text{X}$ for $r >$ 0.7 kpc.


Comparison with the numerical simulation in \citet{Schneider20} shows the same discrepancy (the last three columns in Table \ref{tab:CompVout}). In their simulations, a bi-conical outflow naturally develops, which resembles M 82. They adopt an SFR = 20 M$_{\odot}$ yr$^{-1}$, so we reduce their predicted densities by a factor 20/8 = 2.5. In their model, the starburst extends to a radius of 1 kpc, so we only compare their predicted range in outflow density to our data at radii of 1 and 2 kpc. The predicted pressures are 30 to 70 times lower than our measurements.

We note that in both the model and simulation, the rapid radial drop in the predicted density in the warm gas is caused by the rapid drop in the thermal pressure ($P_\text{th}$) in the wind fluid (via both a $r^{-2}$ drop in density and by the associated adiabatic cooling), and by the assumption that the warm ionized clouds are in balance with $P_\text{th}$ of the wind fluid. This mismatch between the models and the data implies that the clouds we observed are highly over-pressured relative to $P_\text{th}$ of the wind fluid. In contrast, \cite{Heckman90} and \cite{Lehnert96} showed that the radial density profiles in starburst outflows (including M 82) could be explained if the cloud pressure is set by the ram pressure ($P_\text{ram}$) of the wind fluid. The ratio of the ram and thermal pressure in the wind fluid will be 5/3 $M^{2}$, where $M$ is the Mach number in the wind fluid. Since $M \gg 1$ in these supersonic winds, the pressure differences can be substantial.

To compute the ram pressure on the clouds, we need to use the relative velocity between the wind and the cloud rather than the wind velocity. Doing so, we then find that in both the \citet{Fielding22} models and \citet{Schneider20} simulations, the ram pressures on the clouds are indeed significantly larger than the thermal pressures, and in better agreement with the M 82 data (see \text{the right panel in Figure \ref{fig:CompModelFB2}} and Table \ref{tab:CompVout}). 

We note that existing numerical simulations of wind-cloud interactions are not consistent with a balance between cloud thermal pressure and wind ram pressure. Regardless of the physical basis of the disagreement between theory and the data, this implies that both the numerical simulations and the semi-analytic model are missing some important physics. It is then unclear how this missing physics would affect the other theoretical predictions.

\subsubsection{Other Cloud Properties}

The results presented in \citet{Fielding22} allow us to compare other key properties of their clouds to what we have estimated for M 82, namely the cloud radii and column densities (i.e., \Rcloud, and \NHcloud\ derived in Section \ref{sec:structure}). 



\citet{Fielding22} do not plot the modelled \Rcloud, but these can be inferred from their \Mcloud\ and \NHcloud\ values. At a fiducial distance of 1 kpc from the starburst the implied \Rcloud\ range from $\sim$ 5 to 200 pc for \Mcloud\ = 10$^1$ to 10$^6$ M$_{\odot}$, respectively. Our measured \Rcloud\ in M 82 ($<$ 0.9 pc) are less than the lowest mass clouds in the models.

Then we can compare \NHcloud\ between the models and the data. At a fiducial distance of 1 kpc,  we find a range in the models of $N_H$ from $1 \times 10^{19}$ to 8 $\times 10^{20}$ cm$^{-2}$ for the clouds with masses of 10$^1$ to 10$^6$ M$_{\odot}$, respectively. Here, the observed M 82 warm ionized clouds are quite similar to the values in the models (see Figure \ref{fig:OutflowClouds}).

We conclude that the cloud in the models have similar column density but are too large compared with our observed warm ionized outflows for M 82. In the future, we will apply similar models to the CLASSY sample to study this discrepancy more generally.




\section{Conclusion}
\label{sec:conclusion}

In this paper, we have reported the first estimates of the radial distributions of the gas density, outflow rates and cloud properties for the warm ionized gas in the M 82 wind based on the rest-optical data from the Subaru telescope. The main results are summarized as follows:

\begin{itemize}
    \item We have derived the radial distribution of outflow densities based on [\sii] \ly6717, 6731 emission lines. We find that the density drops from $\sim$ 200 cm$^{-3}$ at $r$ = 0.5 kpc to $\sim$ 40 cm$^{-3}$ at $r$ = 2.2 kpc, while the best-fit power-law is $\ne(r) = 100 \times (\frac{r}{1165 pc})^{-1.17}$ (Figure \ref{fig:OutflowDensity} and Section \ref{sec:density}).

    \item We calculated the radial distribution of the lateral width of the outflow based on the Subaru/FOCAS image of M 82. We find that the lateral width is $\sim$ 1.5 $\times$ $r$, where $r$ is the outflow distance to the galactic center (Section \ref{sec:width}). This leads to a total solid angle of 0.8 $\pi$ ster for the bi-conical \ha\ outflow in M 82.

    \item Based on the derived outflow densities and widths, we then estimated the radial distributions of the volume filling factor (FF), which drops from $\sim$ 10$^{-3}$ to 10$^{-4}$ over the range of $r =$ 0.5 to 2.2 kpc. This leads to the best-fit power-law as FF$(r) = 10^{-3} \times (\frac{r}{628 pc})^{-1.8}$.
    
    \item We measured the mass/energy/momentum outflow rates and found that they drop quite slowly with radius, and stay almost unchanged between 0.8 and 2.2 kpc (Section \ref{sec:rate}). This suggests that the galactic winds in M 82 can indeed supply mass,  momentum, and kinetic energy from the central regions out to at least a few kpc with minimal losses.
   
    \item We compared our derived maps of outflow rates from the warm ionized gas traced by \ha\ to the ones from other outflowing phases in the literature. We found that the cold atomic gas traced by \hi\ 21 cm emissions yields similar mass outflow rates compared with the warm ionized gas, but the latter carries substantially more momentum and kinetic energy. Additionally, comparing with the warm ionized gas, we found that the cold molecular gas traced by the CO emission line yields too larger mass outflow rates, but smaller momentum and kinetic energy outflow rates. In both cases, the differences are because the outflow velocity detected in \ha\ is 3 -- 4 times larger than the velocities seen in the \hi\ 21 cm and CO emission lines. These velocities are consistent with a picture in which the atomic and molecular gas actually trace a fountain flow extending out to a few kpc (Section \ref{sec:CompOutflowRates}).

    \item By comparing to a large sample of local star-forming galaxies in the CLASSY sample studied using UV absorption-lines, we found the warm ionized outflow probed by the \ha\ emission lines in M 82 follows similar scaling relationships. This suggests that the outflow properties in M 82 are similar to other local star-forming and starburst galaxies. The consistency between CLASSY and M 82 also suggests that the ionized gas seen in emission and absorption is likely to trace similar material (Section \ref{sec:CompareCLASSY}).

     \item We estimated the radial distributions of various outflow cloud properties, including sizes and column densities (Section \ref{sec:structure}). We found the the cloud sizes are typically 0.1 -- 0.9 pc and cloud column densities are 10$^{19.1}$ -- 10$^{20.7}$ cm$^{-2}$, which drops steadily from 0.5 to 2.0 kpc. These values are 0.2 -- 2 dex smaller than the ones measured for the clouds in the warm ionized outflows based on the UV absorption-line data for the CLASSY sample.

    \item We compared the warm ionized outflows in M 82 with the theoretical models and simulations from \cite{Schneider20} and \cite{Fielding22} in Section \ref{sec:CompTheory}. After accounting for geometrical and SFR differences, we found that the thermal pressures in the clouds predicted by these models are far smaller than our measured valuesin the M 82 data. There is better agreement with the wind ram pressures in the models and simulations. 


    Overall, we have presented novel measurements of radial distributions of outflow properties in M 82, including the outflow velocity, density, rates, and cloud properties. 
    Our work motivates similar spatially resolved studies of the ionized gas in a larger sample of galactic winds. Various essential questions await answers. For example, what are the general radial distributions of outflow properties and associated feedback? How can we adopt these radial distributions to help constrain future models and simulations of galactic winds? How are the different phases of winds connected spatially and kinetically, based on high spatial resolution, multi-wavelength data? Answering these questions will ultimately reveal the complex properties and structures of galactic winds and resolve their feedback on their host galaxies.
    
\end{itemize}

\begin{acknowledgements}

This research is based on data collected at the Subaru Telescope, which is operated by the National Astronomical Observatory of Japan. We are honored and grateful for the opportunity of observing the Universe from Maunakea, which has the cultural, historical, and natural significance in Hawaii. 

X.X. and T.H. thank D. Fielding, G. Bryan, and M. Gronke for useful discussions.


\end{acknowledgements}

\facilities{Subaru Telescope}

\typeout{} 
\bibliography{main}{}
\bibliographystyle{aasjournal}



\end{document}